# Automated identification of current sheets − a new tool to study turbulence and intermittency in the solar wind

Olga Khabarova[1,2], Timothy Sagitov[3], Roman Kislov[1,2], and Gang Li[4]

[1] Pushkov Institute of Terrestrial Magnetism, Ionosphere and Radiowave Propagation of the Russian Academy of Sciences (IZMIRAN), Troitsk, Moscow 108840, Russia; habarova@izmiran.ru .

[2] Space Research Institute of the Russian Academy of Sciences (IKI), Moscow 117997, Russia.

[3] Higher School of Economics University (HSE), Moscow 101000, Russia

[4] Center for Space Plasma and Aeronomic Research (CSPAR), University of Alabama in Huntsville, Huntsville, AL 35805, USA.

Corresponding author: Olga Khabarova (habarova@izmiran.ru)

**Key Points:**

- A new method of the automated identification of current sheets is proposed and the current sheet database https://csdb.izmiran.ru is created.
- The number of CSs per day (R) is determined by variations of the kinetic and thermal energy density of the solar wind.
- R dramatically increases in stream/corotating interaction regions and interplanetary mass ejection sheaths.

**Abstract**

We propose a new method of the automated identification of current sheets (CSs) that represents a formalization of the visual inspection approach employed in case studies. CSs are often identified by eye via the analysis of characteristic changes in the interplanetary magnetic field (IMF) and plasma parameters. Known visual and semi-automated empirical methods of CS identification are exact but do not allow a comprehensive statistical analysis of CS properties. Existing automated methods partially solve this problem. Meanwhile, these methods suggest an analysis of variations of the IMF and its direction only. In our three-parameter empirical method, we employ both the solar wind plasma and IMF parameters to identify CSs of various types. Derivatives of the IMF strength, the plasma beta and the ratio of the Alfvén speed $V_A$ to the solar wind speed $V$ taken with the one-second cadence are used. We find that the CS daily rate $R$ correlates with the solar wind temperature $T$ rather than with $V$ and is proportional to the sum of the kinetic and thermal energy density $\sim V^2(N+5N')+10T(N+N')$, where $N'=2$cm$^{-3}$ is the background level of the solar wind density $N$. Maxima of $R$ are associated with stream/corotating interaction regions and interplanetary mass ejection sheaths. A multiyear list of CSs identified at 1 AU can be found at https://csdb.izmiran.ru .

**Plain Language Summary**

We formalize an experience of observers in identifying current sheets (CS) via the analysis of typical changes in the interplanetary magnetic field and solar wind plasma parameters. A new automated method of CS identification is created, and an open access multi-year database of CSs observed at 1 AU is compiled (see https://csdb.izmiran.ru). We find that the daily rate of CSs is determined by variations of the kinetic and thermal energy density of the solar wind.

**1 Introduction**

Current sheets (CSs) in the solar wind are specific discontinuities carrying the electric current that contributes to the global electric circuit of the heliosphere (Wilcox & Ness 1965, Svalgaard et al., 1983; Suess et al., 2009; Kislov et al., 2015, 2019; Maiewski et al., 2020). On the one hand, these structures may be formed as a result of turbulence and various dynamical processes occurring in the solar wind, and, on the other hand, some CSs are of solar origin. One of the most important features known about CSs is that, independently of their origin, their width is approximately the same (~ several proton gyroradii), while their elongation varies considerably depending on the way of their formation (Malova et al., 2017). The CSs that represent elongated neutral lines of the solar magnetic field are the most stable structures that may extend to the outer heliosphere, and, in turn, the turbulence-born CSs are much shorter, very dynamic and unstable by nature (Li, 2008; Zhdankin et al., 2013; Podesta, 2017; Zelenyi et al., 2020).

The last decade brought a lot of discoveries about properties of CSs in space plasmas, mainly owing to the fast development of the numerical analysis methods such as magnetohydrodynamic (MHD) numerical simulations of fluids that allow studying CS formation in turbulent plasmas (e.g., www.mhdturbulence.com; Burkhart et al., 2020), the Test Particle (TP) and Particle-In-Cell (PIC) coding on supercomputers (e.g., Hesse et al., 2001; Pritchett, 2003; Muñoz & Büchner 2018; Xia & Zharkova, 2018, 2020), also implementing General-Purpose Graphics Processing Units (GPGPU) (e.g.,

https://gyires.inf.unideb.hu/KMITT/a53/ch05.html ; Dokken et al., 2007; Mingalev et al., 2019, 2020). Theoretical studies of processes associated with CSs, including numerical simulations employing hybrid models, PIC and MHD/Hall-MHD simulations, show that CSs effectively contribute to the solar wind dynamics and energy transformation occurring at different scales, from MHD energy-containing scales to kinetic (see Servidio et al. 2010; Donato et al. 2012; Zhdankin et al. 2013; Wan et al. 2015; Papini et al. 2019; Pezzi et al. 2021 and references therein).

Recent advances in case studies of CSs in the solar wind have also contributed to better understanding of the CS fine structure and processes associated with CS dynamics. Overall, there were confirmed (i) 3-D stochastic or turbulent nature of CSs, (ii) self-organization of CSs, (iii) their multi-layered fine structure, (iv) a strong connection between CSs and magnetic islands/plasmoids, and (v) their important role in particle acceleration in space plasmas (e.g., Zharkova and Khabarova, 2012, 2015; Eriksson et al. 2014; Khabarova et al., 2015, 2016, 2021; Khabarova & Zank, 2017; le Roux et al. 2019; Adhikari et al., 2019; Malandraki et al., 2019; Lazarian et al. 2020; Tan, 2020). Li et al. (2011) and, recently, Borovsky & Burkholder (2020) have shown from observations that the presence of CSs in the solar wind determines a power spectral index of high-frequency magnetic field variations, presumably, not only because of CS typical scales and frequency of their detection by a spacecraft but also due to various dynamical processes occurring at CSs. This is consistent with the results of numerical simulations of Franci et al. (2017) who have found that magnetic reconnection at ion-scale CSs leads to the development of the turbulent cascade at sub-ion scales, impacting spectra of magnetic fluctuations. In turn, it is known from numerical testing that turbulence enhances magnetic reconnection at current sheets (see Kowal et al. 2009 and references therein) and it also leads to the formation of new CSs (e.g., Cerri & Califano, 2017). Therefore, since CSs are structures mutually linked with magnetic reconnection and turbulence, it is very important to know their properties as well as properties of the plasma containing them.

CSs in the solar wind are associated with two types of turbulence. First, strong and long-lived CSs mainly of the solar origin represent a source of secondary CSs and 3-D flux ropes/plasmoids (or 2-D magnetic islands) created by instabilities and magnetic reconnection that undergoes especially intensively if such CSs are disturbed. Dynamical processes occurring at strong CSs lead to the formation of a wide analogue of the heliospheric plasma sheet around them. This region is usually treated as turbulent or having signatures of intermittency. On the other hand, turbulent solar wind far from such CSs is never too quiet and always shows signatures of turbulent cascade leading to the occurrence of small-scale and short-lived CSs and magnetic islands observed from the Sun to the outer heliosphere (see Khabarova et al. 2021; Pezzi et al. 2021 and references therein).

In contrast to case studies and simulations, statistical studies of CSs are infrequent in this area (e.g., Li, 2008; Zhang et al., 2008; Suess et al., 2009; Borovsky & Denton, 2011; Li et al., 2011; Malova et al., 2017; Podesta, 2017), and, obviously, a lot of information is still missed because of the lack of a comprehensive statistical analysis. The absence of an easily accessible database of CSs subsequently identified for a prolonged period lays researches under a necessity to compile their own short CS lists coming from case studies (e.g., Suess et al., 2009; Malova et al., 2017; Burkholder & Otto, 2019; Borovsky & Burkholder, 2020; https://lasp.colorado.edu/mms/sdc/public/about/events/#/. The magnetic reconnection exhaust list compiled by J. Gosling (http://www.srl.caltech.edu/ACE/ASC/DATA/level3/swepam/ACE_ExhaustList.pdf) may also be considered as a list of manually-picked CSs appropriate for statistical purposes since reconnection exhausts are located in the nearest vicinity of corresponding reconnecting current sheets.

Such lists typically consist of tens-hundreds CSs or reconnection exhaust crossings randomly picked for months or even years, while the rate of the occurrence of CSs is ~ hundreds CSs per hour (see Li (2008), Podesta (2017), and results of this study). Therefore, a problem of the insufficient progress in CS studies is that the manual identification of CSs remains the most frequently used way to analyze CS properties. A significant loss of information under such an approach is obvious.

To solve the problem, methods of the CS identification that suggest some automatization have been proposed. An automated method of identification of discontinuities of various types can be used first, - for example, a Partial Variance of Increments (PVI) method (Greco et al., 2009, 2018) that identifies all discontinuities, including shocks, or a method of the identification of magnetic holes carrying current sheets at their borders (Winterhalter et al., 1994; Zhang et al., 2008). If the identification of coherent structures is subsequently complemented by a visual inspection, this allows extracting of CSs from the whole body of events (e.g., Zhang et al., 2008; Malandraki et al., 2019). An additional automated check for features typical for current sheets (for example, an analysis of the magnetic field shear angle variations) is possible too (e.g., Yordanova et al., 2020). Such methods are called semi-automated.

There are fully automated methods allowing the CS identification (Li, 2008; Zhdankin et al., 2013; Podesta, 2017; Azizabadi et al., 2020; Pecora et al., 2021). The most popular among them are Gang Li's method (Li, 2008) and John Podesta's method (Podesta, 2017). Gang Li's method is based on the analysis of the IMF angle variations, supposing that a CS crossing is always accompanied by a sharp change in the local magnetic field direction (Li, 2008; Li et al., 2011; Miao et al., 2011). The method works well in turbulent plasmas. The current density calculation is the other way to find a CS location (Podesta, 2017). To calculate the necessary derivatives, one should perform a transition from time dependences **B**(t) to spatial **B**($\xi$). When studying CSs associated with turbulence, Taylor's hypothesis is usually employed, according to which disturbances propagate together with the plasma flow, i.e. d**B**/dt – $V_{SW}$ d**B**/d$\xi$ = 0, where $V_{SW}$ is the solar wind speed, **B** is the magnetic field vector, and $\xi$ is the coordinate along the solar wind speed direction (see Podesta, 2017 for detailed explanations). Using Taylor's hypothesis, one can identify CSs embedded in specific flows and consisting of particles that move with the structure. Within this approach, all automated methods cited above can successfully be applied to the turbulent solar wind.

Meanwhile, some CSs are associated with plasma structures which propagate with the speed different from the surrounding solar wind speed, i.e. such current-carrying structures are not embedded in the freely expanding solar wind and consist of different particles all the time. This is the case of quasi-stationary CSs formed (i) at shocks and strong discontinuities, including the heliospheric current sheet (HCS) and other CSs of the solar origin, the terrestrial magnetopause, leading edges of interplanetary coronal mass ejections (ICMEs), and fast flows from coronal holes, and (ii) owing to interactions of wave fronts (Wilcox & Ness, 1965; Svalgaard et al., 1983; Dunlop et al., 2002; Khabarova et al., 2015, 2016, 2017a, b, 2018; Khabarova & Zank, 2017; Le Roux et al. 2019; Malandraki et al., 2019). If a spacecraft crosses such a structure, an attempt to calculate the electric current density according to Taylor's hypothesis may lead to an error, and the obtained electric current may have an incorrect magnitude and direction. Furthermore, the larger-scale CSs of the non-turbulent origin may create or be surrounded by other structures, namely, secondary CSs and magnetic islands that cannot formally be attributed to turbulence either but is treated as intermittent structures in case studies (e.g., Khabarova et al., 2015, Adhikari et al., 2019). A notable example is the heliospheric plasma sheet (HPS) surrounding the HCS (Winterhalter et al., 1994; Simunac et al., 2012). Multiple structures observed within the HPS represent both CSs originated from the extension of coronal streamers and locally-born coherent structures, namely, CSs and magnetic islands (Khabarova & Zastenker, 2011; Khabarova et al., 2015, 2017b, 2018; Maiewski et al., 2020).

Therefore, it makes sense to enhance a CS identification based on variations of the magnetic field by considering additional plasma features widely used by specialists to find CSs by eye (Suess et al., 2009; Khabarova et al., 2015, 2016, 2017b, 2018, 2020; Khabarova & Zank 2017; Malova et al., 2017; Adhikari et al., 2019). Note that the method proposed in this work is not based on calculating the current density and is independent of the applicability/inapplicability of Taylor's hypothesis. Meanwhile, we discuss it in Section 4 in connection with validation of our method that suggests calculation of the electric current density. In this work, we formalize the experience of observers in visual inspection of typical changes of the IMF and the solar wind plasma parameters at CS crossings, turning it into a new automated method of CS identification. A multi-year list of CSs observed at 1 AU is compiled as a result of this study. The creation of an open-access database allows a new level of statistical analysis to be applied to a large number of current sheets since their properties may sometimes be missed or incorrectly understood from the analysis of small samples of current sheets. The database is also useful for case studies since observers can instantly find a location of current sheets within a time interval under study. Finally, knowing the number of current sheets per interval and their properties, one may study turbulence and intermittency in different samples of the solar wind.

## 2 Data and Method

### 2.1 Data

Our study is mainly based on an analysis of the Advanced Composition Explorer (ACE) measurements. ACE is one of the key 1 AU spacecraft located between the Earth and the Sun at the 1st Lagrange point (see http://www.srl.caltech.edu/ACE/ , Stone et al., 1998). We have used the following ACE data for 2004-2010: the IMF vector **B** measured with the one second resolution, the corresponding <|**B**|> (the magnetic field magnitude), and the 64-second-resolution plasma data, namely, the solar wind proton number density $N$, the solar wind bulk speed $V$, and the radial component of the proton temperature $T$ (https://cdaweb.gsfc.nasa.gov/)

The other dataset employed is from the STEREO Ahead (A) and STEREO Behind (B) spacecraft that move nearly along the Earth's orbit in the opposite direction with respect to each other (see https://stereo-ssc.nascom.nasa.gov/data.shtml and Kaiser et al., 2008). STEREO A is a little closer to the Sun, and STEREO B is a little further from the Sun than the Earth (Kaiser et al., 2008). The STEREO mission data on $B$ with the one second resolution https://stereo-dev.epss.ucla.edu/l1_data , and the plasma parameters ($N$, $V$ and $T$) with the 60 second resolution are obtained from https://cdaweb.gsfc.nasa.gov/. Currently, only STEREO A remains operating, but we use the STEREO A and STEREO B data for several-day-long intervals taken in 2007-2010 when both spacecraft functioned normally. The STEREO data are used in validation of our method but not employed in the current sheet identification for creating the open-access database.

The plasma data have been interpolated and calculated with the one-second span that allows us to compute the plasma beta $\beta$ (the ratio of the plasma pressure to the magnetic pressure) and the Alfvén speed $V_A$ with the span corresponding to the IMF resolution (see the parameter derivation technique at https://omniweb.gsfc.nasa.gov/ftpbrowser/bow_derivation.html ). At the next step, the one-second derivatives of $B$, $\beta$, and $V_A/V$ are calculated to identify CSs (see the Method).

In order to validate our method, we identify a stream interaction region (SIR) from ACE and STEREO A data and trace it, following 3-D pictures of the normalized density in the ecliptic plane reconstructed with the ENLIL modeling that show the solar wind in white-light as if obtained from the Heliospheric Imager (HI) instruments on board of the STEREO spacecraft (http://helioweather.net ). Using HIs is a unique way to look at the dynamic picture of the behavior of streams/flows in the interplanetary medium. One can find key information about HI observations in white light, the techniques allowing the solar wind density reconstruction from the observations, and the related studies in the following articles: Bisi et al. (2008); Eyles et al. (2008, 2009); Rouillard et al. (2008); Jackson et al. (2009); Scott et al. (2019); Barnard et al. (2020). Here, the ENLIL HI density reconstructions are used to check if the SIR identified by us in situ with two spacecraft is an uninterrupted and freely propagating/rotating structure. The latter is necessary to validate our method with an alternative technique.

Finally, a CS database for ACE compiled by Gang Li according to his method of automated CS identification (Li, 2008) has been compared with the database resulting from our study (https://csdb.izmiran.ru).

### 2.2 Method

We suggest a formalization of the long-time experience of observers in the visual identification of CSs based on the analysis of the IMF and plasma parameters that vary sharply at CSs of different origins in the solar wind (Behannon et al., 1981; Blanco et al., 2006; Zhang et al., 2008; Suess et al., 2009; Simunac et al., 2012; Zharkova & Khabarova, 2012, 2015; Khabarova et al., 2015, 2016; Khabarova & Zank 2017; Malova at al. 2017; Adhikari et al., 2019). As an example, a characteristic crossing of a CS detected by the Wind spacecraft at 1 AU on 25 June 2004 is shown in Figure 1 (modified from Khabarova et al., 2021). The IMF magnitude $B$ sharply decreases because of the neutral line crossing. $B_x$, the IMF component in the Earth-Sun (x) direction in the GSE coordinate system, and the other in-ecliptic component, $B_y$, pass through zero at the CS. Generally, a CS crossing is characterized by a sign reverse of at least one of the IMF components (two in the particular case). Consequently, the azimuthal IMF angle $B_{phi}$ sharply varies, indicating the IMF vector direction changes to the opposite. $V$ and $T$ may slightly increase at CSs, owing to ongoing magnetic reconnection (Gosling et al., 2005; Adhikari et al., 2019; Phan et al., 2020). In the particular case, there are clear signatures of magnetic reconnection at the CS reflected in +/- spikes in the $Vy$ component encompassing the CS and an increase in the temperature at the CS (not shown). The solar wind density increases at the CS, which, in combination with the decreasing $B$ leads to the plasma beta increase at the CS crossing.

The solar wind speed change itself is usually too weak and cannot be used for the CS identification. A decrease of the $V_A/V$ ratio observed at CSs is usually substantial and may be considered as an important key to recognize CSs. Figure 2 shows statistical results obtained by Suess et al. 2009 for CSs identified in 2004–2006 and early 2007. The solar wind proton density increases in the nearest vicinity of CSs, and the $V_A/V_p$ ratio statistically decreases at CSs.

It should be noted that although CSs are crossed for seconds, their crossings are seen differently under different time/space resolutions. Crossings of strong CSs (the HCS, for example) statistically analyzed with the hourly and/or daily resolution are associated with an increase in $B$. This is a common signature of the approach of an observational device to any electric current-carrying surface/conductor, as known in all scientific branches, including geophysics, since it comes from the Biot-Savart law. A bell-like profile of the total magnetic field is typical for crossings of such structures. In the solar wind, the electric current flows in CSs in a form of the net of thin wires (see, e.g. Lazarian et al., (2012) and Kowal et al., (2012)), which makes even small-scale profiles of specific CSs a bit complicated and depending on the way they are crossed by a spacecraft. Magnetic reconnection and formation of plasmoids at CSs also result in a complex behavior of key solar wind parameters in a wider vicinity of such CSs. However, at the statistical level, the |B| increase near strong CSs is clearly seen on large scales. For example, a superposed epoch analysis shows that daily averaged $B$ peaks on the day of the HCS crossing (Khabarova & Zastenker, 2011), while under at least a minute resolution, the HCS crossing is characterized by the decrease in $B$.

The IMF strength and density increase observed around the HCS at large scales may also be explained by the occurrence of dynamical processes at CSs that change properties of plasma surrounding CSs (see Zharkova & Khabarova, 2012, 2015; Zelenyi et al., 2016; Malova et al., 2017, 2018 and references therein), while the IMF strength decrease at the midplane of strong CSs observed with a high resolution is simply explained by the occurrence of the zero B line at which at least one of the magnetic field components crosses zero (Zimbardo et al., 2004; Malova et al. 2017, 2018). Here we are talking about observations of CSs with a high resolution.

The main features seen with a resolution not worse than one minute that may characterize a CS crossing are following: (i) a decrease in $B$, (ii) a decrease in $V_A/V$, and (iii) an increase in $\beta$. It is easy to find that all other peculiarities are linked with the listed ones. We develop a combined method based on that employed by observers since current sheet crossings may be reflected differently in changes of the solar wind/IMF parameters, i.e. sometimes not all of the signatures mentioned above appear altogether. If one wants not to miss as many current sheets as possible, it is better to consider a combination of different signatures.

Since the automatization of the CS recognition process requires setting the same rules for CSs occurring in different plasmas under different conditions, normalization should be performed. Hence, after obtaining $B$, $V_A/V$, and $\beta$, we calculate their one-second derivatives. One can suggest then that spikes of the derivatives reflect the location of CSs.

Noise cutoff or a threshold choice is the most sensitive point of all methods for identifying current sheets. Its introduction is always based on experience and common sense. We have chosen to carefully compare the results with clear cases of CS crossings discussed in literature and find the optimal parameter levels above which the peaks correspond to the CS location in the best way. The first threshold of $\beta=3$ for a one second resolution data has been imposed. The next important point is that the signatures of a CS crossing discussed above should be considered altogether. In statistical terms this means finding a maximum correlation between different datasets to restrict final results by those complying with necessary signatures observed in all key parameters. Applying the maximum correlation condition, we finally obtain the following thresholds: values above $dB/dt= -0.14$, below $d\beta/dt=0.11$, and above $d(V_A/V)/dt= -0.003$ are considered as noise; here $t = 1$ s. Note that the thresholds will be different if one considers a lower resolution or averaged data. We treat variations in *dB/dt* as the main feature, therefore only the spikes that appear simultaneously in *dB/dt* and any of two other parameters are considered as pointing out the CS location.

Additionally to the list of current sheets obtained with the method, we compute the magnetic field shear angle observed at the identified current sheets using the well-known formula (Li 2008; Miao et al. 2011; Yordanova et al. 2020):

$$d\theta = \arccos\left((\mathbf{B}(t), \mathbf{B}(t+dt))/(|\mathbf{B}(t)|\cdot|\mathbf{B}(t+dt)|)\right) \quad (1)$$

We modify the formula as

$$d\theta = \arccos\left((\mathbf{B}(t-16), \mathbf{B}(t+16))/(|\mathbf{B}(t-16)|\cdot|\mathbf{B}(t+16)|)\right), \quad (2)$$

using the time step dt=16 seconds. These are the key points on which a new method of the automated identification of CSs is based. Details are illustrated below step by step in the process of identification of CSs embedded in a SIR.

## 3 Identification of CSs – results

### 3.1 Illustration of finding CSs from the ACE and STEREO data

To demonstrate how the method works, we apply the technique to a SIR, the region not freely propagating in the solar wind but resulting from the interaction of a rotating high-speed flow from a coronal hole with the ambient slower solar wind. A SIR resembles an ICME sheath by properties since it is equally turbulent and full of numerous discontinuities, CSs and magnetic islands (e.g., Ho et al., 1996; Jian et al., 2006; Tessein et al., 2011; Khabarova et al., 2017b). A description of key features of SIRs and their long-lived counterparts, corotating interaction regions (CIRs), can be found in a comprehensive review by Ian Richardon (see Richardson, 2018 and references therein).

SIRs are important drivers of space weather, living in the interplanetary medium much longer than ICMEs that quickly expand, pass 1 AU in 1-3 days and fade with distance. Since SIRs are linked to coronal hole flows, they rotate for many days, striking planets and spacecraft one by one, which gives us an opportunity to study these large-scale structures as a whole and in detail. SIRs/CIRs can be traced remotely via HIs designed to observe dynamics of dense solar wind structures in white light (see 2.1 Data). Such a tracking is necessary (i) to be sure that two different spacecraft detect the same flow (if one analyzes in situ measurements), (ii) to be confident that the flow is not interrupted by an ICME for time enough to observe it in situ, and, finally, (iii ) to analyze a longitudinal evolution of SIRs/CIRs.

Figure 3 is an example of the analysis of the SIR evolution using both the predicted STEREO HI remote view of the SIR and in situ measurements. Figures 3a and 3b show an ecliptic cut of a reconstructed image of the solar wind density as predicted by ENLIL for STEREO HI. The Earth is the green dot, the Sun is in the center, seen from the solar north pole, and two SIRs resemble rotating sleeves. The intensity of grey corresponds to the normalized density value (darker means denser). The red circle indicates the SIR subsequently detected by ACE and STEREO A with a several-day delay. For the corresponding movie see http://helioweather.net, click Archive – date - anim-sta1dej/anim-stb1dej. We use these reconstructions to be sure that the particular SIR reached both ACE and STEREO A uninterrupted (see Section 4.1 below). The SIR arrival is identified via in situ data from both spacecraft.

A SIR represents a transition area between the fast solar wind associated with a coronal hole and the surrounding slow solar wind. The SIR is characterized by compression reflected in the enhanced $N$ and $B$, which is typically observed at its outer edge, and $V$ mostly begins to increase after the crossing of the stream interface (SI), a tangential discontinuity separating slow and fast solar wind within the SIR (e.g., Richardson 2018), although some signatures of the $V$ growth can sometimes be seen before the SI crossing. SIRs may be imagined as dense turbulent shells surrounding high-speed rotating flows from coronal holes.

Figure 3c shows that a typical feature characterizing an approach of the high-speed coronal hole flow is a growth of $N$ in the background of a constant or slightly increasing $V$. After the passage of the SI at which $N$ sharply falls but still remains above quiet period values and $V$ sharply increases, the spacecraft occurs in the part of the SIR affected by the high-speed coronal hole flow in which $V$ typically keeps growing. The SI located inside the SIR is a very bright and easy-to-identify structure that allows studying the SIR rotation via in situ observations successfully. In our case, we track the SI to compute the electric current density to check the location of CSs by an independent method (see 4. Validation).

The leading part of the SIR marked by yellow in Figure 3c has been used as a test bench to compute derivatives of $B$, $\beta$, and $V_A/V$ (see 2.2 Method). In this particular case, the beginning of the SIR was detected by ACE on 25 June 2010 16:00:50, and the SI within the SIR was crossed on 26 June 2010 02:48:18. The same SIR arrived at STEREO A on 30 June 2010 17:11:49, and the SI was detected on 1 July 2010 03:29:48.

In Figure 4 we show a typical example of the CS identification for the period of the ACE encounter with the SIR. Another example is the identification of CSs in a fragment of the same SIR observed by STEREO A (see Figure 5). Figure 5 is analogous to Figure 4 but shows CSs identified in another time interval. The threshold for cutting off the noise is shown by the horizontal red line for each parameter. The spikes occurring out of the noise level indicate the CS location. The noise level is calculated as described in 2.2 Method. Only the spikes simultaneously observed in the $B$ derivative and any of the other parameters are treated as CS indicators.

As a result of the study, we have compiled a 1 AU CS list with the one second cadence for 1998-2010. The CS database representing a set of the three-month-length CS lists is available at the dedicated IZMIRAN website https://csdb.izmiran.ru . Additionally to indicating a location of each CS, the list contains the following parameters observed at the CS crossing: $B, N, V, T, \beta, V_A$, and three derivatives on the analysis of which the method is based (see 2.2 Method).

So far, the 1 AU current sheet database consists of the three-month-length "output_ACE_1s_YYYY_MM-MM.csv" files and the format "IMF_angle_YYYY_MM-MM" indicate files with computed shear angles at CSs. The current sheet lists are compiled with the one-second cadence. The plain text files contain information about the current sheet location and the corresponding plasma and IMF parameters:

1. date (location of the current sheet pointed with the one-second resolution),
2. <|B|>, nT (the IMF magnitude) - $B$ in the manuscript,
3. H_DENSITY_#/cc  (the solar wind density, 1/cm^3) - $N$ in the manuscript,
4. SW_H_SPEED_km/s (the solar wind proton speed, km/s) - $V$ in the manuscript,
5. H_TEMP_RADIAL_Kelvin (the radial component of the solar wind temperature, K) - $T$ in the manuscript,
6. Beta (the plasma beta $\beta$),
7. VA (the Alfvén speed, km/s) - $V_A$ ,

8. B_der (the derivative of <|B|>) - $dB/dt$,
9. Beta_der (the $\beta$ derivative) - $d\beta/dt$,
10. VA/V_der (the derivative of the Alfvén speed to the solar wind speed ratio) - $d(V_A/V)/dt$ in the manuscript.

This is an open access database, and the website does not require a registration, but a reference to the website and this article is necessary if one uses the CS database and/or the method.

### 3.2 What determines the current sheet occurrence in the solar wind?

It is easy to find in Figure 4 and Figure 5 that the number of CSs may reach hundreds per hour or ~several thousands per day in the most turbulent regions. Figure 6 illustrates this with a CS daily occurrence rate or, in other words, the number of current sheets per day ($R$) as observed in 1998-2010 (Figure 6a). The red solid curve represents $R$ smoothed using a 27-point Savitzky-Golay filter with the 3rd degree polynomial (Savitzky & Golay, 1964), and the dash line is a one-year smoothed $R$. Figure 6b shows the number of sunspots per day smoothed in the same way in order to estimate if $R$ depends on solar cycle. If one calculates a Pearson coefficient of correlation of $R$ vs the sunspot number based on initial daily data from OMNIweb (https://omniweb.gsfc.nasa.gov/ow.html ), the result will be -0.08, which means no correlation at all. Smoothing increases it, as shown in Figure 6c. A 27-day-window smooth gives the correlation coefficient of ~-0.44, which still means no correlation, and a one-year-window gives a negative correlation of ~-0.67. Note that the increase of the correlation coefficient upon increasing the window up to the total length of the time period analyzed is usually considered as an artificial statistical result of smoothing rather than a physical result. A conclusive solar cycle dependence/independence analysis can be performed with a larger database only. So far, we can suggest that, upon Figure 6, there is no obvious solar cycle dependence although some connection between $R$ and the sunspot number is possible due to the $R$ dependence on the SIR and ICME occurrence rates (see a related discussion in Section 5). Further investigations will be made in this area when we extend the database.

The bright feature that meets the eye is quasi-regular variations seen in $R$. We find that this is a reflection of the CS production increase in turbulent and intermittent regions associated with SIRs/CIRs, the ICME sheaths and the related increase in the solar wind energy flow. A preliminary analysis allows us to conclude that the highest peaks of $R$ seen in Figure 6 correspond to either SIR/CIR or ICME sheath observation periods. One can compare known ICME/SIR lists (see http://www.srl.caltech.edu/ACE/ASC/DATA/level3/index.html) with $R$ to see this feature. Using our database, we show this in Figure 7 and Figure 8.

Figure 7 depicts results of a superposed epoch analysis of $R$ versus the day with respect to the ICME sheath/SIR crossing, as observed in 1998-2010. In the SIR case (the upper panel), $R$ begins to grow before the SIR arrival, and its peak occurs at the SI. Since the location of SI varies from one SIR to another, there are several secondary peaks around the main one. The R profile around ICMEs (the lower panel) is different. The number of current sheets per hour increases exactly at the beginning of the ICME sheath crossing, i.e. mostly at the shock, and decreases much faster than observed in the SIR case, which is understandable because SIRs are generally wider than ICME sheaths.

One may suggest that the found peaking of $R$ within ICME sheaths/SIRs is related to the turbulence level observed downstream of ICME-driven interplanetary shocks and within SIRs in which the SI represents the strongest discontinuity at 1 AU. It is quite probable that the bottom figure will change dramatically at farther heliocentric distances since SIR-associated shocks are usually formed farther than 1 AU, and properties of turbulent plasma within SIRs should vary with distance.

The *R* distributions for hours in which SIRs are observed (red bars) and for hours without SIRs (blue bars) can be found in Figure 8a, and Figure 8b shows histograms of the distribution of R observed within ICME sheaths (red bars) and outside ICME sheaths (blue bars). One can see that red histograms are shifted to larger values in both cases, which confirms that the number of current sheets increases in ICME sheaths and SIRs.

We also illustrate this point with Figure 9 that shows typical variations in *R* observed before, during and after a SIR (Figure 9a) and an ICME (Figure 9b). The SIR in Figure 9a was observed in the period of 3-5 May 2004. This is SIR no. 1 shown in Figure 3 of Jian et al., (2011). A shock pair did not form at the SIR's edges at 1 AU. Figure 9a shows that *R* quickly reaches its maximum within the SIR and slowly decreases afterwards within the associated coronal hole flow.

The very strong ICME detected by ACE from 15 May 2005 to 17 May 2005 was characterized by a classic ICME-driven forward interplanetary shock followed by the compressed turbulent sheath (see Figure 1 of Dasso et al., 2009) with which the *R* peak is associated (Figure 9b). An important feature seen in Figure 9b is a precursor representing a prolonged moderate *R* increase before the ICME approach. This phenomenon has been discussed before in terms of the crossing of a magnetic cavity filled with magnetic islands/plasmoids or flux ropes formed between the HCS and an ICME (Khabarova et al., 2016, 2017b, 2018; Adhikari et al., 2019). Note that the increase of the number of magnetic islands and the increase of the CS number are linked since magnetic islands are separated by CSs (e.g., Malandraki et al., 2019).

*R* correlates with *T* in a higher degree than with *V* and other key plasma/IMF parameters. The corresponding Pearson correlation coefficients $C_{R-X}$ ($X = T, V, B, N$) calculated for the entire database from 2004 to 2010 are as follows: $C_{R-T}= 0.66$, $C_{R-V}= 0.46$, $C_{R-B}= 0.43$, and $C_{R-N}= 0.21$.

Let us derive a formula that expresses *R* as a function of physical parameters. CSs are characterized by an increased energy density compared to the surrounding plasma. In particular, CSs can be distinguished by the extrema of *β* and *B* because of that. Therefore, one can expect the number of CSs per day to be proportional to the average energy density in such regions. We will search for the best correlation between *R* and key solar wind parameters, trying to find a function (i.e., the best-fit parameter $R_E$) that depends on the kinetic energy density (the first term), the thermal energy density (the second term), and the magnetic field energy density (the third term).

$$R_E = \alpha_1 m_p V^2(N + a_2)/2 + \alpha_3 (N + \alpha_4) k T + \alpha_5 B^2. \qquad (3)$$

Here $\alpha_{1,2,3,4,5}$ are dimensional constant factors, $m_p$ the proton mass, and *k* denotes the Boltzmann constant. *V* is measured in km/s, *T* - in K, *N* - in cm$^{-3}$, and *B* is measured in nT. Parameters $\alpha_1$, $\alpha_3$, and $\alpha_5$ describe the impact of the corresponding types of energy on the CS rate, while $\alpha_2$ and $\alpha_4$ reflect the presence of the low-density plasma, in the background of which $R_E$ variations caused by SIRs/CIRs and ICMEs occur.

An empirical search for the best-fit parameter shows that observed *R* highly correlates with a function of the following form:

$$R_E = \left(V^2 (N+5\, N') + 10( N+ N')T\right)/5000 \qquad (4)$$

Here, all values are reduced to the units of (1), and $N' =2$ cm$^{-3}$ corresponds to the concentration observed at 1 AU in the background solar wind. The magnetic field energy term is absent in (4) since its inclusion does not lead to obtaining a better result. The basis of a composite function method of the best correlation finding is described in Khabarova & Zastenker (2011) and Khabarova & Savin (2015). The normalization factor of 1/5000 is chosen to make $R_E \approx R$.

Figure 10 shows examples of the comparison of $R$ vs $T$ and $R$ vs the best-fit parameter in the form of (4) as observed for three months in 2004 (Figure 10a) and 2010 (Figure 10b). The correlation coefficient between $R$ and the best-fit parameter is 0.82 for the whole 2004-2010 database. It may reach ~0.9 in some months. In particular, the correlation between $R$ and the best-fit parameter is 0.87 for the period shown in Figure 10a and 0.86 for the period shown in Figure 10b. This is far larger than any of the correlation coefficients $C_{R-X}$ ($X=T, V, B, N$) calculated for the same period.

One can find that (4) has the following general form in the CGS metric system:

$$R_E = \left(1.65 \cdot 10^{-2} \, (\rho + 5\rho') \, V^2 / 2 + 10 \, (N + N') \, kT \right)/(5000 \, k \, cm^{-3} \, K) = R. \qquad (5)$$

Here $\rho = m_p N$ is the solar wind density, and $\rho' = m_p N'$. The denominator in (5) is a constant dimensional normalization factor. Formula (5) is useful for a theoretical analysis. Meanwhile, observers will obtain $R$ directly from the empirical formula (4), taking the parameters from the ACE database, since $R_E \approx R$.

## 4 Validation of the method

### 4.1 Validation by the electric current density calculation

In order to check if the method of identifying CSs proposed above reflects reality in terms of revealing spatial variations of the electric current, we will estimate the current density **j** and compare the location of its peaks with the location of CSs identified with our method.

To calculate **j** in a chosen SIR/CIR's region, one should know details of the angular rotation of the SIR/CIR, which can be found if the SIR/CIR is subsequently detected by one spacecraft after another. First,

$$\text{rot}\mathbf{B} = \mu_0 \mathbf{j} \, , \qquad (6)$$

here $\mu_0 = 1.256637062 \cdot 10^{-6}$ H/m in the SI system of units. The electric displacement currents can be ignored in the case of slowly evolving CSs. To have (6) in the form of the dependence of **B** on coordinates, it is necessary to understand how a particular CIR/SIR moves. Here, we assume for simplicity that the only movement is CIR/SIR's rotation around the rotation axis of the Sun in the ecliptic plane with the angular velocity ω in the chosen reference Radial-Tangential-Normal frame (RTN).

In speculations and calculations below, r denotes the radial coordinate in the RTN reference system, the normal component is S, and the tangential is W. Then, $r = r_{AU} \cos(\omega t + \varphi)$ and $W = r_{AU} \sin(\omega t + \varphi)$, where $\varphi$ is the initial phase, $r_{AU}$ is the distance from the Sun to the spacecraft, which is different for each spacecraft involved in calculations.

Considering **B**(t) as **B**(t(r, W, S)), where t is time, it is easy to find that

$$\mu_0 \, j_N = - \, dB_W/dt \, \sin(\omega t + \varphi)/\omega r_{AU} - dB_r/dt \, \cos(\omega t + \varphi)/\omega r_{AU}. \qquad (7)$$

The initial phase φ can be chosen arbitrarily. However, it determines whether the first and second terms are of the same order or not, which potentially may lead to the diminishing of the role of one of the magnetic field components. Therefore, it is reasonable to take the following quantity $j_0$ to estimate the current density:

$$\mu_0 \, j_0 = - \, (dB_W/dt)/\omega r_{AU} - (dB_r/dt)/\omega r_{AU}. \qquad (8)$$

Here, $j_0$ is the upper limit for the electric current density in the S direction.

The next step is the calculation of the angular velocity. Let us introduce the following notations:

- $\Omega_L$ is the angular velocity of the spacecraft with respect to the Sun, where L = 1, 2 denotes, respectively, the 1st and the 2nd spacecraft that subsequently detect the rotating CIR/SIR front;
- $\Omega$ is the angular velocity of the CIR/SIR with respect to the Sun, determined by the difference $t$ of the moments of arrival of the CIR/SIR front to the 1st ($t_1$) and 2nd spacecraft ($t_2$);
- $HCI_{L,1}$ is the heliographic longitude of the spacecraft L at $t_1$ in the Heliocentric Inertial (HCI) system of coordinates, and
- $HCI_{L,2}$ is the heliographic longitude of the spacecraft L at the point of time $t_2$, in HCI.

Then the corresponding angular velocities are: $\Omega_1 = (HCI_{1,2} - HCI_{1,1}) / (t_2 - t_1)$ ; $\Omega_2 = (HCI_{2,2} - HCI_{2,1}) / (t_2 - t_1)$ ; $\Omega = (HCI_{2,2} - HCI_{1,1}) / (t_2 - t_1)$. The angular velocities of each spacecraft are different, so the angular velocity of the CIR/SIR relative to each spacecraft will be different. Therefore, $\omega = \Omega - \Omega_L$ is the angular velocity of the CIR/SIR with respect to the spacecraft L=1, 2.

The final step is the derivation of the formula for the electric current density. For the convenience of calculations, we transform (6) to the following form:

$j_0 = -w \ ( \ dB_W / dt + dB_r / dt \ )$ \hspace{2em} (9),

where

$w = (\mu_0 \ r_{AU} \ \omega)^{-1} = 10^4 / (1.2566 \cdot 7 \cdot 215 \cdot [\omega] \cdot [r_{AU}])$ \hspace{1em} (10)

and $[\omega]$ is the angular velocity of the CIR/SIR with respect to the spacecraft under consideration, multiplied by $10^6$. $[r_{AU}]$ is the position of the spacecraft with respect to the Sun in astronomical units. If the magnetic field is measured in nT, then the current density will be in $nA/m^2$.

Note that since only sharp changes in the current density but not the absolute values matter for the analysis we propose, both $[\omega]$ and $[r_{AU}]$ can be considered as constants in SI for simplification. Meanwhile, other tasks would require detailed calculations of the parameters as described above. This technique can be considered as an additional and independent method of CS identification.

Figure 11a and Figure 11b represent a pair of two-panel graphs, the upper of which is for the electric current density and the lower depict the localization of CSs derived from the three-parameter method. In the upper panel, we define current sheets as peaks with a height exceeding 0.5 $nA/m^2$, thus cutting off the background currents. For easier comparison, a typical "0-1" view of the CS location is modulated by the electric current density corresponding to each time at which a CS is identified (see the lower panels in Figure 11). Figure 11a shows calculations based on the ACE data, and Figure 11b shows the same for STEREO A. Fragments are arbitrarily chosen from the "yellow stripe" region indicated in Figure 3. When calculating the correlation between the two rows, we take the current density from the validation method: if it is higher than the background the value is 1, otherwise it is 0. Comparing the location of spikes of the electric current density with the location of CSs from our CS database, one can conclude that the correlation between the two is very high, reaching 0.9.

4.2 Validation by the comparison with Gang Li's method of CS identifying

Gang Li's method (Li, 2008) is based on the analysis of local variations of the IMF direction that become dramatic at CSs. It is usually applied to identifying CSs in turbulent plasmas (see Introduction). On the one hand, we understand that the three-parameter method is completely different from (Li, 2008), but on the other hand, it would be useful to compare results of both.

We apply both techniques to the ACE data, analyzing the location of current sheets observed in 2004. Figure 12a and Figure 12b show two fragments of rows of CSs identified by our method (the upper panels) and Li's method (the lower panels). The obtained results are promising. Despite the differences mentioned above, the correlation between the two rows is ~0.8. The fact that two methods give little different pictures of the current sheet location is understandable. The mismatches seen in Figure 12 may be determined (i) by different targets of two methods (sometimes they identify not overlapping classes of CSs) and/or (ii) by different sensitivity of the methods because of the different cut off levels (see Introduction and Conclusions). A detailed comparison of different methods identifying current sheets is a future task. Here, it is important to find that the two methods show similar results.

## 5 Discussion and Conclusions

A new method of the automated identification of current sheets has been created. We use a simple approach popular among observers to find CSs via a visual analysis of the behavior of both IMF and plasma parameters. The method suggests identifying CSs of various types, from short and unstable turbulence-born CSs embedded in the streaming solar wind to quasi-stable CSs associated with wave processes and large-scale solar-connected objects. Note that our method is not aimed at identifying all current sheets in the solar wind but at creating a database that might be used for comprehensive statistical studies. A list of CSs identified at 1 AU from the 1998-2012 ACE data can be found at https://csdb.izmiran.ru, and the list will be expanded soon.

The main statistical results obtained with the CS database are as follows:
- On average, one-three thousand CSs are detected daily at the Earth's orbit. The number of CSs per day is determined by the sum of the kinetic and thermal energy densities in a high degree (the correlation coefficient is ~0.8).

The best-fit parameter is $(V^2(N+5\,N') + 10(N+N')T)/5000$ if $V$ is given in km/s, $T$ - in K, $N$ is given in cm$^{-3}$, $B$ is measured in nT, and $N'=2$ cm$^{-3}$. The second, thermal term makes a one order larger input in the parameter than the kinetic one since the correlation with the temperature is better than with the speed. Meanwhile, without the first term, it is impossible to follow smaller-scale variations in the CS daily rate.
- The daily CS occurrence is found to be almost insensitive to the magnetic energy density variations.
- Peaks of the number of CSs per day are found to occur in SIRs/CIRs and ICME sheaths. The result is confirmed statistically.
- There is clustering of CSs. This fact is in agreement with the facts that (i) in realistic plasmas, the electric current is predicted to flow along multiple CSs rather than along one (Kowal et al., 2012; Lazarian et al. 2020), (ii) multiple CSs with the electric current of the same direction tend to merge, and, on the other hand, (iii) strong CSs create an analogue of the plasma sheet consisting of secondary CSs around (e.g., Malova et al., 2017).
- There is no obvious connection between the daily CS rate and the solar cycle. However, this preliminary conclusion should be reconsidered after the expansion of the CS database to several solar cycles and carrying out an analysis of the nature of long-term variations of the CS rate.

It should be noted that an impact of the solar cycle on the daily CS rate is possible. The presence of a negative correlation between $R$ and the sunspot number at large temporal scales may indicate that the behavior of $R$ is slightly impacted by ICMEs and mostly determined by SIRs. Indeed, the occurrence rate of ICMEs follows the solar cycle very well, correlating positively, (Möstl et al. 2020) while the SIR occurrence rate is more irregular and generally anti-correlates with the sunspot number (Zhang et al. 2007; Yermolaev et al. 2012). The variable relative impact of the rates may lead to the observed picture of $R$ variations over the 11-year solar cycle. A further study will be carried out to clarify this point.

Using the created database, it is possible to study properties of both CSs (at small scales) and the solar wind containing them (at larger scales). Since CSs are associated with turbulent plasmas, their rate may be considered as one of characteristics of turbulence and intermittency (as shown in Malandraki et al. 2019). For example, effects of clustering and an increase of the number of current sheets in SIRs/CIRs and ICME sheaths definitely reflect properties of turbulence in the corresponding regions. A turbulent ICME sheath created downstream of the interplanetary shock is a well-known source of CSs and magnetic islands (see Chian and Muñoz, 2011; Zank et al., 2015; Khabarova et al., 2021; Pezzi et al. 2021 and references therein). Less is known about the formation of CSs in SIRs (Khabarova et al., 2021) although SIRs are treated as turbulent regions as well (Richardson 2018). CSs also represent separators of magnetic islands occurring in turbulent plasma (Khabarova et al. 2015, 2016, 2021; Zank et al., 2015; Le Roux et al. 2019; Malandraki et al. 2019), therefore it would be perspective to combine the database of current sheets (https://csdb.izmiran.ru) and the database of flux ropes http://fluxrope.info (Zheng and Hu, 2018), studying associated effects, including particle acceleration in the areas filled with CSs and magnetic islands.

Understanding the physical nature of the obtained dependence between the CS occurrence and the solar wind energy density is a subject of future studies since it may be interpreted in different ways. On the one hand, a good correlation between the CS occurrence and the solar wind temperature has been known for years. It is often explained in terms of the solar wind heating by dissipation at turbulence-born CSs (e.g., Karimabadi et al., 2013; Wu et al., 2013; Wan et al., 2015, 2016; Adhikari et al., 2017). Meanwhile, the heating effect is supposed to be localized in the nearest vicinity of CSs, which is debated, especially for strong CSs. For example, Borovsky & Denton (2011) claim no correlation between strong CSs and local heating.

One can suggest from our results that CSs are intensively formed in hot plasma flows rather than represent a direct source of heating. Indeed, in addition to turbulence, CSs can appear at plasma irregularities caused by thermal fluctuations and asymmetries, which increase in regions with the raised temperature (Landau & Lifshitz, 1980, Lifshitz & Pitaevskii, 1981). CSs occur in such regions owing to the development of instabilities in plasma with temperature anisotropies and the increased magnetic reconnection rate (see Gingell et al., 2015 and references therein). Therefore, CSs may be formed due to local processes caused by large-scale plasma motions like those associated with SIRs/CIRs and ICMEs.

Substantial heating of particles can also be determined by intermittency, when the impact of turbulence-associated effects mentioned above is not essential. An intermittency phenomenon takes place if there are alternating regions with ordered laminar and turbulent flows (Landau and Lifshitz 1987, Landau et al., 1984). Regular quasi-stationary structures separated from chaotic regions by CSs are associated with such flows. In this case, the CSs can potentially gather in "sandwiches" consisting of several CSs due to the spatial quasi-periodicity (e.g., Bykov et al., 2008). This is consistent with the observed clustering of CSs. The sandwich-like regions may scatter some particles passing through them and accelerate other particles, the phase of which turns out to be favorable upon entering the layer, for example, in the way suggested by Zelenyi et al. (2011).

The other aspect is that one may suggest that the good correlation between the CS occurrence rate and $T$ is just because of the good $T$-$V$ correlation (Borovsky & Denton, 2011). However, (i) the CS daily rate does not correlate with $V$ well, and (ii) the good $T$-$V$ correlation paradigm cannot be applied to SIRs and ICMEs in which $R$ reaches its peak values, as seen from observations. Using an ICME as an example, Matthaeus et al. (2006) have shown that the correlation between $V$ and $T$ is significantly reduced in the presence of non-spherically-symmetric processes. One can see an illustration of this effect in Figure 6. $T$ increases only for a short period at the leading edges of the fast speed coronal hole flow (Figure 6a) and the ICME (Figure 6b), but $V$ remains high many hours after that. More studies are necessary to clarify this point.

Overall, many questions regarding properties of CSs in the solar wind and their impact on plasma heating and particle acceleration remain open. The CS list compiled as a result of this study opens an opportunity to answer the questions. The database will grow and include data for the entire period of 1 AU in situ observations starting with IMP8 and ending with the DSCOVR spacecraft. We also plan to compile CS lists for the STEREO spacecraft, Ulysses, Parker Solar Probe and Solar Orbiter. The database is open access, and the community members are welcome to employ the method and the CS list for their statistical and case studies.

**Acknowledgments and Data**

We are grateful to the CDAWeb team for providing open access data for the magnetic field from ACE (N. Ness, Bartol Research Institute) and the ACE/SWEPAM Solar Wind Experiment 64-second level 2 data (D.J. McComas, SWRI) at the CDAWeb platform https://cdaweb.gsfc.nasa.gov . The STEREO mission data are obtained from the UCLA STEREO Data Server (https://stereo-dev.epss.ucla.edu/l1_data) and CDAWeb (https://cdaweb.gsfc.nasa.gov).

We thank Dusan Odstrcil for providing the solar wind density reconstructions from STEREO HI on the ENLIL Solar Wind Prediction website http://helioweather.net . The sunspot number data are obtained from the OMNIWeb website https://omniweb.gsfc.nasa.gov/ow.html , thanks to Dr. Natalia Papitashvili and Robert Candey.

The database resulting from our study is available at https://csdb.izmiran.ru . The authors thank Andrei Osin for his help in maintaining the website at the IZMIRAN institutional server.

O.K. and R.K. are supported by Russian Science Foundation grant No. 20-42-04418. T.S. acknowledges the HSE's general support and encouragement of student's scientific activity.

**References**

Adhikari, L., G.P. Zank, P. Hunana, D. Shiota, R. Bruno, Q. Hu, and D. Telloni (2017), II. Transport of nearly incompressible magnetohydrodynamic turbulence from 1 to 75 au, *Astrophys. J.*, 841, 85, aa6f5d, doi: 10.3847/1538-4357/aa6f5d

Adhikari, L., O. Khabarova, G.P. Zank, and L.-L. Zhao (2019), The Role of Magnetic Reconnection–associated Processes in Local Particle Acceleration in the Solar Wind, *Astrophys. J.*, 873, 1, id. 72, doi: 10.3847/1538-4357/ab05c6

Azizabadi, A.C., N. Jain, and J. Büchner (2020), Identification and characterization of current sheets in collisionless plasma turbulence, arXiv:2009.03881, https://arxiv.org/pdf/2009.03881

Barnard, L., M.J. Owens, C.J. Scott, and C.A. de Koning (2020), Ensemble CME modeling constrained by heliospheric imager observations, *AGU Advances*, **1**, e2020AV000214. doi: 10.1029/2020AV000214

Behannon, K.W., F.M. Neubauer, and H. Barnstorf (1981), Fine scale characteristics of interplanetary sector boundaries. *J. Geophys. Res.*, 86, 3273, doi: 10.1029/JA086iA05p03273

Bisi, M. M., B. V. Jackson, P. P. Hick, A. Buffington, D. Odstrcil, and J. M. Clover (2008), Three-dimensional reconstructions of the early November 2004 Coordinated Data Analysis Workshop geomagnetic storms: Analyses of STELab IPS speed and SMEI density data, *J. Geophys. Res.*, 113, A00A11, doi:10.1029/2008JA013222

Blanco J.J., J. Rodriguez-Pacheco, M.A. Hidalgo, and J. Sequeiros (2006), Analysis of the heliospheric current sheet fine structure: Single or multiple current sheets, *J. Atm. Sol. Terr. Phys.*, 68, 2173-2181, doi: 10.1016/j.jastp.2006.08.007

Borovsky, J. E., and B.L. Burkholder (2020), On the Fourier contribution of strong current sheets to the high-frequency magnetic power spectral density of the solar wind, *J. Geophys. Res.*, 125, e2019JA027307, https://doi.org/10.1029/2019JA027307

Borovsky, J. E., and M.H. Denton (2011), No evidence for heating of the solar wind at strong current sheets, *Astrophys. J. Lett.*, 739, L61, https://doi.org/10.1088/2041-8205/739/2/L61

Burkhart, B., S.M. Appel, S. Bialy, J. Cho, A.J. Christensen, D. Collins, C. Federrath, D.B. Fielding, D. Finkbeiner, and A.S. Hill (2020), *Astrophys. J.*, 905, abc484, doi: 10.3847/1538-4357/abc484

Burkholder, B.L., and A. Otto (2019), Magnetic reconnection of solar flux tubes and coronal reconnection signatures in the solar wind at 1 AU, J. Geophys. Res., 124, 8227–8254, doi:10.1029/2019JA027114

Bykov, A.A., L.M. Zelenyi, and K.V. Malova (2008), Triple splitting of a thin current sheet: A new type of plasma equilibrium, *Plasma Phys. Rep.*, 34, 128–134, doi: 10.1134/S1063780X08020050

Cerri, S.S., and F. Califano (2017), Reconnection and small-scale fields in 2D-3V hybrid-kinetic driven turbulence simulations, *New J. Phys.*, 19, 025007, doi: 10.1088/1367-2630/aa5c4a

Chian, A. C.-L., and P.R. Muñoz (2011), Detection of Current Sheets and Magnetic Reconnections at the Turbulent Leading Edge of an Interplanetary Coronal Mass Ejection, *Astrophys. J. Lett.*, 733, L34, doi:10.1088/2041-8205/733/2/L34

Conlon, T.M., S.E. Milan, J.A. Davies, and A.O. Williams (2015), Corotating Interaction Regions as Seen by the STEREO Heliospheric Imagers 2007 – 2010, *Sol. Phys.*, 290, 2291–2309, doi: 10.1007/s11207-015-0759-z

Dasso, S., C.H. Mandrini, B. Schmieder, H. Cremades, C. Cid, Y. Cerrato, E. Saiz, P. De´moulin, A. N. Zhukov, L. Rodriguez, A. Aran, M. Menvielle, and S. Poedts (2009), Linking two consecutive nonmerging magnetic clouds with their solar sources, J. Geophys. Res., 114, A02109, doi:10.1029/2008JA013102

Dokken, T., T.R. Hagen, and J.M. Hjelmervik (2007), An Introduction to General-Purpose Computing on Programmable Graphics Hardware. In: Hasle G., Lie KA., Quak E. (eds) *Geometric Modelling, Numerical Simulation, and Optimization*. Springer, Berlin, Heidelberg, doi: 10.1007/978-3-540-68783-2_5

Donato S., S. Servidio, P. Dmitruk, V. Carbone, M.A. Shay, P.A. Cassak, and W.H. Matthaeus (2012), Reconnection events in two-dimensional Hall magnetohydrodynamic turbulence. *Phys. Plasmas*, 19, 9, 092307, doi : 10.1063/1.4754151

Eriksson S., D.L. Newman, G. Lapenta, and V. Angelopoulos (2014), On the signatures of magnetic islands and multiple X-lines in the solar wind as observed by ARTEMIS and WIND, *Plasma Phys. Control. Fusion*, 56, 064008, doi : 10.1088/0741-3335/56/6/064008

Eyles, C.J., Harrison, R.A., Davis, C.J. et al. (2009), The Heliospheric Imagers Onboard the STEREO Mission, Sol. Phys., 254, 387–445, https://doi.org/10.1007/s11207-008-9299-0

Eyles, C.J., R.A. Harrison, C.J. Davis et al. (2009), The Heliospheric Imagers Onboard the STEREO Mission, Sol. Phys., 254, 387–445, doi:10.1007/s11207-008-9299-0

Franci, L., S.S. Cerri, F. Califano, S. Landi, E. Papini, A. Verdini, L. Matteini, F. Jenko, and P. Hellinger (2017), Magnetic Reconnection as a Driver for a Sub-ion-scale Cascade in Plasma Turbulence, *Astrophys. J. Lett*., 850, 1, L16, https://doi.org/10.3847/2041-8213/aa93fb

Gingell, P.W., D. Burgess, and L. Matteini (2015), The three-dimensional evolution of ion-scale current sheets: tearing and drift-kink instabilities in the presence of proton temperature anisotropy, *Astrophys. J.*, 802, 4, doi: 10.1088/0004-637X/802/1/4

Gosling, J.T., R.M. Skoug, D.J. McComas, and C.W. Smith (2005), Direct evidence for magnetic reconnection in the solar wind near 1 AU, *J. Geophys. Res*., 110, A01107, doi:10.1029/2004JA010809

Greco, A., W.H. Matthaeus, S. Perri, K.T. Osman, S. Servidio, M. Wan, and P. Dmitruk (2018), Partial Variance of Increments Method in Solar Wind Observations and Plasma Simulations, *Space Sci. Rev*., 214, 1, doi: 10.1007/s11214-017-0435-8

Greco, A., W.H. Matthaeus, S. Servidio, P. Chuychai, and P. Dmitruk (2009), Statistical analysis of discontinuities in solar wind ace data and comparison with intermittent mhd turbulence, *Astrophys J.*, 691, L111, doi: 10.1088/0004-637X/691/2/L111

Hesse, M., M. Kuznetsova, and J. Birn (2001), Particle-in-cell simulations of three-dimensional collisionless magnetic reconnection, *J. Geophys. Res*., 106, 29831-29841, doi: 10.1029/2001JA000075

Ho, C. M., B.T. Tsurutani, R. Sakurai, B.E. Goldstein, A. Balogh, and J.L. Phillips (1996), Interplanetary discontinuities in corotating streams and their interaction regions. *Astron. and Astrophys.*, 316, 346-349, https://ui.adsabs.harvard.edu/#abs/1996A&A...316..346H/abstract

Jackson, B.V., P.P. Hick, A. Buffington, M.M. Bisi, and J.M. Clover (2009), SMEI direct, 3-D-reconstruction sky maps, and volumetric analyses, and their comparison with SOHO and STEREO observations, 27, 4097–4104, *Ann. Geophys*., doi: 10.5194/angeo-27-4097-2009

Jian, L., C.T. Russell, J.G. Luhmann, and R.M. Skoug (2006), Properties of Stream Interactions at One AU During 1995–2004, *Sol. Phys*, 239, 337–392, doi: 10.1007/s11207-006-0132-3

Jian, L.K., C.T. Russell, J.G. Luhmann, P.J. MacNeice, D. Odstrcil, P. Riley, J.A. Linker, R M. Skoug, and J.T. Steinberg (2011), Comparison of Observations at ACE and Ulysses with Enlil Model Results: Stream Interaction Regions During Carrington Rotations 2016 – 2018, *Sol. Phys.*, 273, 179–203, doi: 10.1007/s11207-011-9858-7

Kaiser, M.L., T.A. Kucera, J.M. Davila, O.C.St. Cyr, M. Guhathakurta, and E. Christian (2008), The STEREO Mission: An Introduction, *Space Sci. Rev.* 136, 5–16, doi: 10.1007/s11214-007-9277-0

Karimabadi, H., V. Roytershteyn, W. Daughton, and Y.-H Liu (2013), "Recent Evolution in the Theory of Magnetic Reconnection and Its Connection with Turbulence", Space Sci. Rev., 178, 307–323, doi:10.1007/s11214-013-0021-7

Khabarova O., V. Zharkova, Q. Xia, and O.E. Malandraki (2020), Counterstreaming strahls and heat flux dropouts as possible signatures of local particle acceleration in the solar wind, 894, L12, *Astrophys. J.* Lett., https://doi.org/10.3847/2041-8213/ab8cb8

Khabarova O.V., H.V. Malova, R.A. Kislov, L.M. Zelenyi, V.N. Obridko, A.F. Kharshiladze, M. Tokumaru, J.M. Sokół, S. Grzedzielski, and K. Fujiki (2017a), High-latitude conic current sheets in the solar wind, *Astrophys. J.*, 836, 108, https://doi.org/10.3847/1538-4357/836/1/108

Khabarova O.V., O.E. Malandraki, G.P. Zank, G. Li, J.A. le Roux, and G.M. Webb (2018), Re-Acceleration of Energetic Particles in Large-Scale Heliospheric Magnetic Cavities, *Proc. of the International Astronomical Union*, Symposium S335: Space Weather of the Heliosphere: Processes and Forecasts, July 2017, 13, 75-81, doi: 10.1017/S1743921318000285

Khabarova, O., and G. Zastenker (2011), Sharp changes in solar wind ion flux and density within and out of current sheets, *Sol. Phys.*, 270, 311-329, DOI: 10.1007/s11207-011-9719-4

Khabarova, O., and I. Savin (2015), Changes in Environmental Parameters and Their Impact on Forest Growth in Northern Eurasia, *Atmosph. and Climate Sci.*, 5, 91-105, doi: 10.4236/acs.2015.52007

Khabarova, O., G.P. Zank, G. Li, J.A. le Roux, G.M. Webb, A. Dosch, and O.E. Malandraki (2015), Small-scale magnetic islands in the solar wind and their role in particle acceleration. 1. Dynamics of magnetic islands near the heliospheric current sheet, *Astrophys. J.*, 808, 181, doi: 10.1088/0004-637X/808/2/181

Khabarova, O., G.P. Zank, G. Li, O.E. Malandraki, J.A. le Roux, and G.M. Webb (2016), Small-scale magnetic islands in the solar wind and their role in particle acceleration. II. Particle energization inside magnetically confined cavities, *Astrophys J.*, 827, 122, https://doi.org/10.3847/0004-637X/827/2/122

Khabarova, O., O. Malandraki, H. Malova, et al. (2021), Current Sheets, Plasmoids and Flux Ropes in the Heliosphere. Part I. 2-D or not 2-D? General and Observational Aspects, *Space. Sci. Rev.*, 217, 38, https://doi.org/10.1007/s11214-021-00814-x

Khabarova, O.V., and G.P. Zank (2017), Energetic Particles of keV–MeV Energies Observed near Reconnecting Current Sheets at 1 au, *Astrophys. J.*, 843, 4, https://doi.org/10.3847/1538-4357/aa7686

Khabarova, O.V., G.P. Zank, O.E. Malandraki, G. Li, J.A. le Roux, and G.M. Webb (2017b), Observational Evidence for Local Particle Acceleration Associated with Magnetically Confined Magnetic Islands in the Heliosphere - a Review, *Sun and Geosph.*, 12, 23-30, http://adsabs.harvard.edu/abs/2017SunGe..12...23K

Kislov, R.A., O. Khabarova, and H.V. Malova (2015), A new stationary analytical model of the heliospheric current sheet and the plasma sheet, 120, 8210-8228, *J. Geophys. Res.*, https://doi.org/10.1002/2015JA021294

Kislov, R.A., O.V. Khabarova, and H.V. Malova (2019), Quasi-stationary Current Sheets of the Solar Origin in the Heliosphere, Astrophys J., 875, 28, https://doi.org/10.3847/1538-4357/ab0dff

Kowal, G., A. Lazarian, E.T. Vishniac, and K. Otmianowska-Mazur (2012), Reconnection studies under different types of turbulence driving, *Nonlin. Processes Geophys.*, 19, 297–314, https://doi.org/10.5194/npg-19-297-2012

Kowal, G., A. Lazarian, E.T. Vishniac, and K. Otmianowska-Mazur, (2009), Numerical tests of fast reconnection in weakly stochastic magnetic fields, *Astrophys. J.*, 700, 1, https://doi.org/10.1088/0004-637X/700/1/63

Landau, L. D., & Lifshitz, E. M. (1980). Fluctuations. In L. D. Landau, E. M. Lifshitz (eds.), *Statistical physics (Third edition), Course of theoretical physics* (Vol. 5, pp. 333-400), Oxford: Butterworth-Heinemann, doi: 10.1016/B978-0-08-057046-4.50019-1

Landau, L. D., & Lifshitz, E. M. (1987). Turbulence. In L. D. Landau, E. M. Lifshitz (eds.), *Fluid mechanics (Second edition)*, *Course of theoretical physics* (Vol. 6, pp. 95-156), Oxford: Butterworth-Heinemann, doi: 10.1016/B978-0-08-033933-7.50011-8

Landau, L. D., Lifshitz, E. M., & Pitaevskii, L. P. (1984). Magnetohydrodynamics. In L. D. Landau, E. M. Lifshitz (eds.), *Electrodynamics of continuous media (Second edition)*, *Course of theoretical physics* (Vol. 8, pp. 225-256), Oxford: Butterworth-Heinemann

Lazarian, A., G.L. Eyink, A. Jafari, G. Kowal, H. Li, S. Xu, and E.T. Vishniac (2020), 3D turbulent reconnection: Theory, tests, and astrophysical implications, *Phys. Plasmas*, 27, 012305, doi: 10.1063/1.5110603

Lazarian, A., L. Vlahos, G. Kowal, H. Yan, A. Beresnyak, and E.M. de Gouveia Dal Pino (2012), Turbulence, magnetic reconnection in turbulent fluids and energetic particle acceleration, *Space Sci. Rev.*, 173, 557-622, doi:10.1007/s11214-012-9936-7

le Roux J.A., G.M. Webb, O.V. Khabarova, K.T. Van Eck, L.-L. Zhao, L. Adhikari, Journal of Physics: Conference Series (19th Annual International Astrophysics Conference 9-13 March 2020 in Santa Fe, New Mexico, USA) (2020), 1620, 012008, doi:10.1088/1742-6596/1620/1/012008

Le Roux, J. A., Webb G.M., Khabarova O.V., Zhao L.-L., and Adhikari L. (2019), Modeling Energetic Particle Acceleration and Transport in a Solar Wind Region with Contracting and Reconnecting Small-scale Flux Ropes at Earth Orbit, *Astrophys. J.*, 887, 77, doi: 10.3847/1538-4357/ab521f

le Roux, J. A., Zank, G. P., & Khabarova, O. V. (2018) Self-consistent Energetic Particle Acceleration by Contracting and Reconnecting Small-scale Flux Ropes: The Governing Equations, *Astrophys. J.*, 864, 158, doi: 10.3847/1538-4357/aad8b3

Li G., (2008), Current-Sheet-like Structures in the Solar Wind, *Astrophys. J. Lett.*, 672, L65, doi: 10.1086/525847

Li, G., B. Miao, Q. Hu, and G. Qin (2011), Effect of Current Sheets on the Solar Wind Magnetic Field Power Spectrum from the Ulysses Observation: From Kraichnan to Kolmogorov Scaling, Phys. Rev. Lett., 106, 125001, doi:10.1103/PhysRevLett.106.125001

Lifshitz, E. M., & Pitaevskii, L. P. (1981). Collisions in plasmas. In E. M. Lifshitz, L. P. Pitaevskii (eds.), *Physical kinetics (First edition)*, *Course of theoretical physics* (Vol. 10, pp. 168-216), Oxford: Butterworth-Heinemann, doi: 10.1016/B978-0-08-057049-5.50009-6

Maiewski, E.V., H.V. Malova, R.A. Kislov, V.Yu. Popov, A.A. Petrukovich, O.V. Khabarova, and L.M. Zelenyi (2020), Formation of Multiple Current Sheets in the Heliospheric Plasma Sheet, *Cosmic Res.*, 58, 411–425, doi: 10.1134/S0010952520060076

Malandraki, O., O. Khabarova, R. Bruno, G.P. Zank, G. Li, B. Jackson, M.M. Bisi, A. Greco, O. Pezzi, W. Matthaeus, A.G. Chasapis, S. Servidio, H. Malova, R. Kislov, F. Effenberger, J. le Roux, Y. Chen, Q. Hu, and E. Engelbrecht (2019), Current sheets, magnetic islands and associated particle acceleration in the solar wind as observed by Ulysses near the ecliptic plane, *Astrophys, J.*, 881,116, doi: 10.3847/1538-4357/ab289a

Malova, H.V., V.Yu. Popov, E.E. Grigorenko, A.A. Petrukovich, D. Delcourt, A.S. Sharma, O.V. Khabarova, and L.M. Zelenyi (2017), Evidence for quasi-adiabatic motion of charged particles in strong current sheets in the solar wind, Astrophys. J., 834, 34, doi:10.3847/1538-4357/834/1/34

Malova, Kh.V., V.Yu. Popov, O.V. Khabarova, E.E. Grigorenko, A.A. Petrukovich, and L.M. Zeleny (2018), Structure of Current Sheets with Quasi-Adiabatic Dynamics of Particles in the Solar Wind, *Cosmic Res.*, 56, 462–470, doi: 10.1134/S0010952518060060

Matthaeus, W.H., H.A. Elliott, and D.J. McComas (2006), Correlation of speed and temperature in the solar wind, *J. Geophys. Res.*, 111, A10103, doi:10.1029/2006JA011636

Miao, B., B. Peng, and G. Li (2011), Current sheets from Ulysses observation, *Ann. Geophys.*, 29, 237-249, doi:10.5194/angeo-29-237-2011

Mingalev, O.V., I.V. Mingalev, H.V. Malova, A.M. Merzlyi, V.S. Mingalev, and O.V. Khabarova (2020), Description of Large-Scale Processes in the Near-Earth Space Plasma, Plasma Phys. Rep., 46, 374–395, doi: 10.1134/S1063780X20030083

Mingalev, O.V., O.V. Khabarova, H.V. Malova, I.V. Mingalev, R.A. Kislov, M.N. Mel'nik, P.V. Setsko, L.M. Zelenyi, and G.P. Zank, Sol. Syst. Res. (2019) doi: 10.1134/S0038094619010064

Möstl, C., A.J. Weiss, R.L. Bailey, M.A. Reiss, T. Amerstorfer, J. Hinterreiter, M. Bauer, S.W. McIntosh, N. Lugaz, and D. Stansby (2020), Prediction of the In Situ Coronal Mass Ejection Rate for Solar Cycle 25: Implications for Parker Solar Probe In Situ Observations, *Astrophys. J.*, 903, 2, doi:10.3847/1538-4357/abb9a1

Muñoz, P.A., and Büchner, J. (2018), Kinetic turbulence in fast three-dimensional collisionless guide-field magnetic reconnection, *Phys. Rev. E*, 98, doi:10.1103/physreve.98.043205

Papini E., L. Franci, S. Landi, A. Verdini, L. Matteini, and P. Hellinger (2019) Can Hall magnetohydrodynamics explain plasma turbulence at sub-ion scales?, *Astrophys. J.*, 870, 1, 52, doi: 10.3847/1538-4357/aaf003

Pecora, F., S. Servidio, A. Greco, and W.H. Matthaeus (2021), Identification of coherent structures in space plasmas: The magnetic helicity-PVI method, *Astron. and Astrophys.*, doi: 10.1051/0004-6361/202039639

Pezzi, O., F. Pecora, J. le Roux et al. (2021), Current Sheets, Plasmoids and Flux Ropes in the Heliosphere. Part II: Theoretical Aspects, *Space Sci. Rev.,* 217, 39, doi: 10.1007/s11214-021-00799-7

Phan, T.D., S.D. Bale, J.P. Eastwood, et al., (2020), Parker Solar Probe In Situ Observations of Magnetic Reconnection Exhausts during Encounter 1. *Astrophys. J. Suppl. Ser.,* 246, 34, doi:10.3847/1538-4365/ab55ee

Podesta, J. J. (2017), The most intense current sheets in the high-speed solar wind near 1 AU, *J. Geophys. Res.*, 122, 2795 –2823, doi:10.1002/2016JA023629

Pritchett P.L. (2003) Particle-in-Cell Simulation of Plasmas— A Tutorial. In: Büchner J., Scholer M., Dum C.T. (eds) *Space Plasma Simulation. Lecture Notes in Physics*, 615, Springer, Berlin, Heidelberg, https://doi.org/10.1007/3-540-36530-3_1

Richardson, I.G. (2018), Solar wind stream interaction regions throughout the heliosphere. *Living Rev. in Solar Phys.*, 15, 1, doi:10.1007/s41116-017-0011-z

Rouillard, A.P., J.A. Davies, R.J. Forsyth, et al., (2008), First imaging of corotating interaction regions using the STEREO spacecraft, *Geophys. Res. Lett.*, 35, L10110, doi: 10.1029/2008GL033767

Savitzky, A., and M.J.E. Golay (1964), Smoothing and Differentiation of Data by Simplified Least Squares Procedures, *Analytical Chemistry*, 36, 1627-1639, doi: 10.1021/ac60214a047

Scott, C.J., M.J. Owens, C.A. de Koning, L.A. Barnard, S.R. Jones, and J. Wilkinson (2019), Using ghost fronts within STEREO Heliospheric Imager data to infer the evolution in longitudinal structure of a coronal mass ejection, *Space Weather*, 17, 539–552, doi: 10.1029/2018SW002093

Servidio S., W.H. Matthaeus, M.A. Shay, P. Dmitruk, P.A. Cassak, and M. Wan (2010), Statistics of magnetic reconnection in two-dimensional magnetohydrodynamic turbulence. *Phys. Plasmas*, 17, 3, 032315, doi: 10.1063/1.3368798

Simunac, K.D.C., A.B. Galvin, C.J. Farrugia, et al., (2012), The Heliospheric Plasma Sheet Observed in situ by Three Spacecraft over Four Solar Rotations, *Solar Phys.*, 281, 423-447, doi: 10.1007/s11207-012-0156-9

Stantchev, G., D. Juba, W. Dorland, and A. Varshney (2009), Using Graphics Processors for High-Performance Computation and Visualization of Plasma Turbulence, *Computing in Sci. and Engineering*, 11, 52-59, doi: 10.1109/MCSE.2009.42

Stone, E.C., A.M. Frandsen, R.A. Mewaldt, E.R. Christian, D. Margolies, J.F. Ormes, and F. Snow (1998), The Advanced Composition Explorer, *Space Sci. Rev.*, 86, 1–22, doi: 10.1023/A:1005082526237

Suess, S.T., Y.-K. Ko, R. von Steiger, and R.L. Moore (2009), Quiescent current sheets in the solar wind and origins of slow wind, *J. Geophys. Res.*, 114, A04103, doi: 10.1029/2008JA013704

Svalgaard, L., J.M Wilcox, and T.L. Duvall (1974), A model combining the polar and the sector structured solar magnetic fields, Sol. Phys., 37, 157–172, doi:10.1007/BF00157852

Tan, L.C. (2020), An Alternative Interpretation of Impulsive SEP Events Occurring on 1999 January 9–10, *Astrophys. J.*, 901, 120, doi: 10.3847/1538-4357/abb086

Tessein, J. A., C.W. Smith, B.J. Vasquez, and R.M. Skoug (2011), Turbulence associated with corotating interaction regions at 1 AU: Inertial and dissipation range magnetic field spectra, *J. Geophys. Res.*, 116, A10104, doi:10.1029/2011JA016647

Wan, M., W. H. Matthaeus, V. Roytershteyn, H. Karimabadi, Parashar T. , Wu P. , and Shay M. (2015), Intermittent Dissipation and Heating in 3D Kinetic Plasma Turbulence, *Phys. Rev. Lett.*, 114, 175002, doi: 10.1103/PhysRevLett.114.175002

Wan, M., W.H. Matthaeus, V. Roytershteyn, T.N. Parashar, P. Wu, and H. Karimabadi (2016), Intermittency, coherent structures and dissipation in plasma turbulence, *Phys. of Plasmas,* 23, 042307, doi: 10.1063/1.4945631

Wilcox, J.M., and N.F. Ness (1965), Quasi-stationary corotating structure in the interplanetary medium, J. Geophys. Res., 70, 23, 5793-5805, doi:10.1029/JZ070i023p05793

Winterhalter, D., M. Neugebauer, B. E. Goldstein, E. J. Smith, S. J. Bame, and A. Balogh (1994), Ulysses field and plasma observations of magnetic holes in the solar wind and their relation to mirror-mode structures, *J. Geophys. Res.*, 99, 23371, doi: 10.1029/94JA01977

Wu, P.; Perri S., Osman K., M. Wan, W.H. Matthaeus, M.A. Shay, M.L. Goldstein, H. Karimabadi, and S. Chapman (2013), Intermittent Heating in Solar Wind and Kinetic Simulations, *Astrophys. J. Lett.*, 763, L30, doi: 10.1088/2041-8205/763/2/L30

Xia, Q., & V. Zharkova (2018), Particle acceleration in coalescent and squashed magnetic islands. I. Test particle approach, *Astron. Astrophys.*, 620, A121, doi: 10.1051/0004-6361/201833599

Xia, Q., & V. Zharkova (2020), Particle acceleration in coalescent and squashed magnetic islands. II. Particle-in-cell approach, *Astron. Astrophys.*, 635, A116, doi: 10.1051/0004-6361/201936420

Yermolaev, Y. I., N. S. Nikolaeva, I. G. Lodkina, and M. Y. Yermolaev (2012), Geoeffectiveness and efficiency ofCIR, sheath, and ICME in generation of magnetic storms, *J. Geophys. Res.*, 117, A00L07, doi:10.1029/2011JA017139

Yordanova E., Z. Vörös, S. Raptis, and T. Karlsson (2020), Current Sheet Statistics in the Magnetosheath, *Front. Astron. Space Sci.*, doi: 10.3389/fspas.2020.00002

Zank G.P., P. Hunana, P. Mostafavi, J.A. le Roux, G. Li, G.M.Webb, O. Khabarova, A. Cummings, E. Stone, and R. Decker (2015), Diffusive shock acceleration and reconnection acceleration processes, *Astrophys. J.*, 814, 137, doi: 10.1088/0004-637X/814/2/137

Zelenyi, L.M., H.V. Malova, E.E. Grigorenko, and V.Y. Popov (2016), Thin current sheets: from the work of Ginzburg and Syrovatskii to the present day, *Physics-Uspekhi*, 59, 1057–1090, doi:10.3367/ufne.2016.09.037923

Zelenyi, L.M., H.V. Malova, E.E. Grigorenko, V.Y. Popov, and E.M. Dubinin (2020), Universal scaling of thin current sheets, *Geophys. Res. Lett.*, 47, e2020GL088422, 1-10, doi: 10.1029/2020GL088422

Zelenyi, L.M., S.D. Rybalko, A.V. Artemyev, A.A. Petrukovich, and G. Zimbardo (2011), Charged particle acceleration by intermittent electromagnetic turbulence, *Geophys. Res. Lett.*, 38, L17110, doi:10.1029/2011GL048983

Zhang, J., I.G. Richardson, D.F. Webb, N. Gopalswamy, E. Huttunen, J.C. Kasper, N.V. Nitta, W. Poomvises, B.J. Thompson, C.-C. Wu, S. Yashiro, and A.N. Zhukov (2007), Solar and interplanetary sources of major geomagnetic storms (Dst≤-100 nT) during 1996–2005, *J. Geophys. Res.*, 112, A10102, doi:10.1029/2007JA012321

Zhang, T.L., C.T. Russell, W. Zambelli, Z. Vörös, C. Wang, J.B. Cao, L.K. Jian, R. J. Strangeway, M. Balikhin, W. Baumjohann, M. Delva, M. Volwerk, and K.-H. Glassmeier (2008), Behavior of current sheets at directional magnetic discontinuities in the solar wind at 0.72 AU, *Geophys. Res. Lett.*, 35, L24102, doi: 10.1029/2008GL036120

Zharkova V., and O. Khabarova (2012), Particle dynamics in the reconnecting heliospheric current sheet: Solar wind data versus three-dimensional particle-in-cell simulations, *Astrophys. J.*, 752, 35, doi: 10.1088/0004-637X/752/1/35

Zharkova V., and O. Khabarova (2015), Additional acceleration of solar-wind particles in current sheets of the heliosphere, *Ann. Geophys.* doi: 10.5194/angeo-33-457-2015

Zhdankin, V., D.A. Uzdensky, J.C. Perez, and S. Boldyrev (2013), Statistical analysis of current sheets in three-dimensional magnetohydrodynamic turbulence, *Astrophys. J.*, 771, 124, doi: 10.1088/0004-637X/771/2/124

Zheng, J., and Q. Hu (2018), Observational evidence for self-generation of small-scale magnetic flux ropes from intermittent solar wind turbulence, *Astrophys. J. Lett.*, 852, L23, doi: 0000-0002-7570-2301

Zimbardo, G., A. Greco, P. Veltri, A.L. Taktakishvili, and L.M. Zelenyi (2004), Double peak structure and diamagnetic wings of the magnetotail current sheet, *Ann. Geo.*, 22, 2541-2546, doi:10.5194/angeo-22-2541-2004

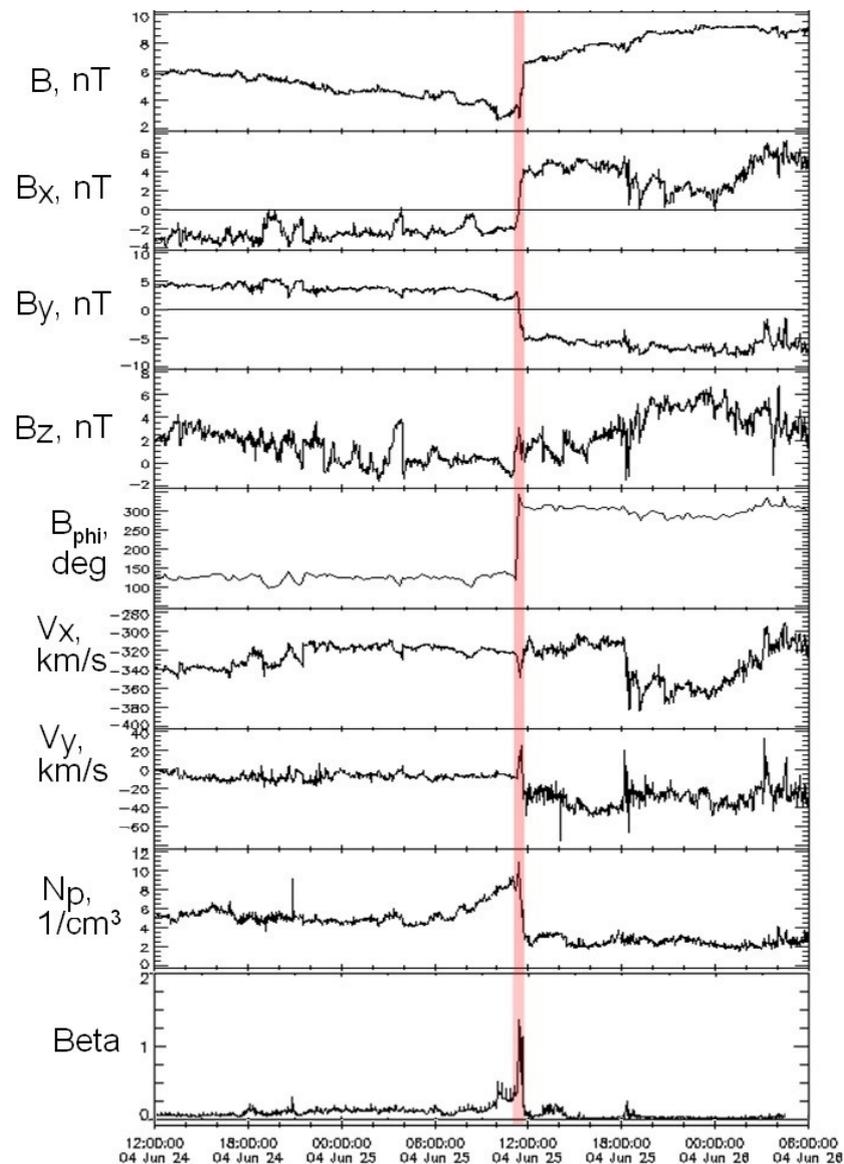

Figure 1. Typical behavior of the IMF and plasma parameters at a CS crossing observed by the Wind spacecraft on 25 June 2004. From top to bottom: the IMF magnitude, the IMF components $B_x$, $B_y$, and $B_z$ (GSE), the IMF azimuthal angle $B_{phi}$, the solar wind speed $V_x$ (the dominant component) and $V_y$ component, the solar wind proton density $Np$, and the plasma beta. CS crossing and the associated changes in the IMF and plasma parameters are depicted by the pink stripe and arrows. Modified from Khabarova et al. (2021).

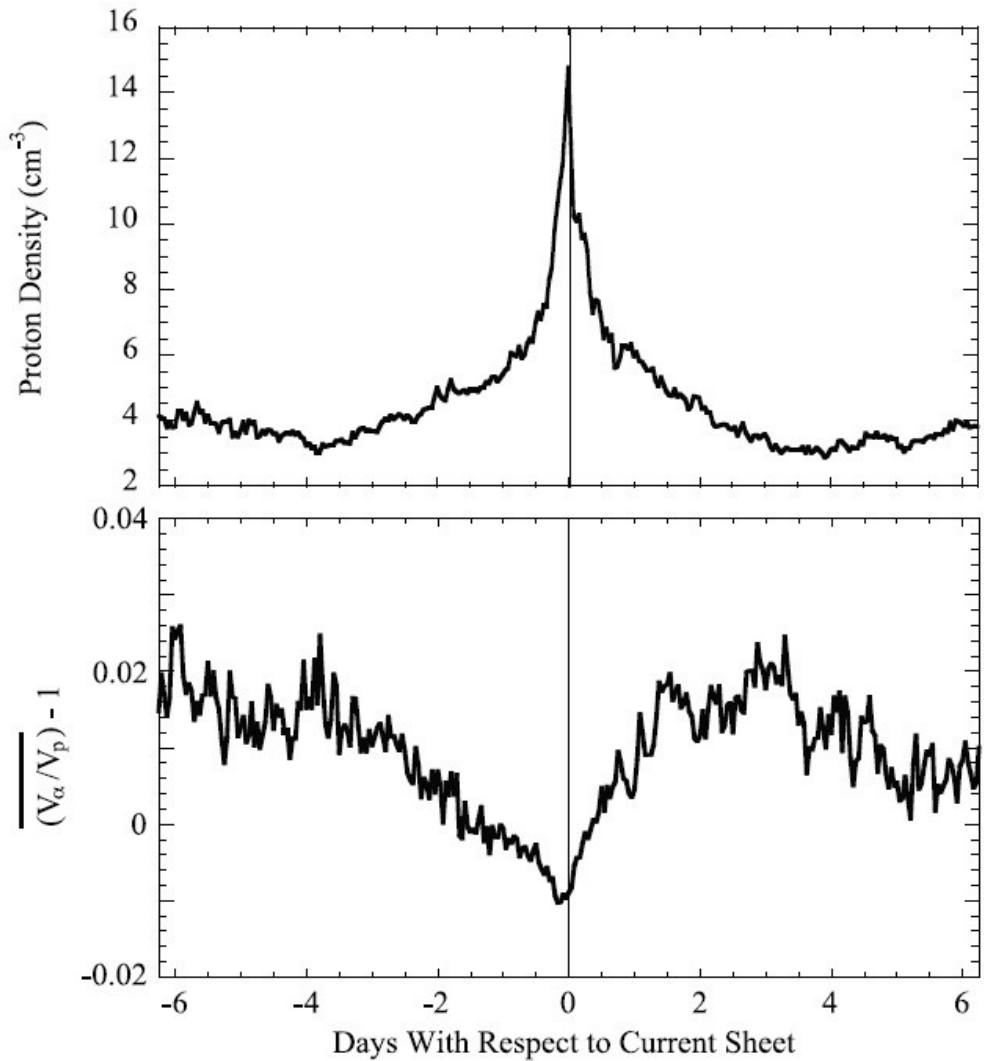

Figure 2. Statistical (superposed epoch analysis) results for the solar wind proton density (upper panel) and the Alfven speed to the solar wind proton speed $V_a/V_p$ ratio in the vicinity of CSs identified by Suess et al. (2009) in 2004–2006 and early 2007. ACE data. Adapted from Suess et al. (2009). Permission to reprint is obtained from the Copyright Clearance Center (License Number 5061650837423).

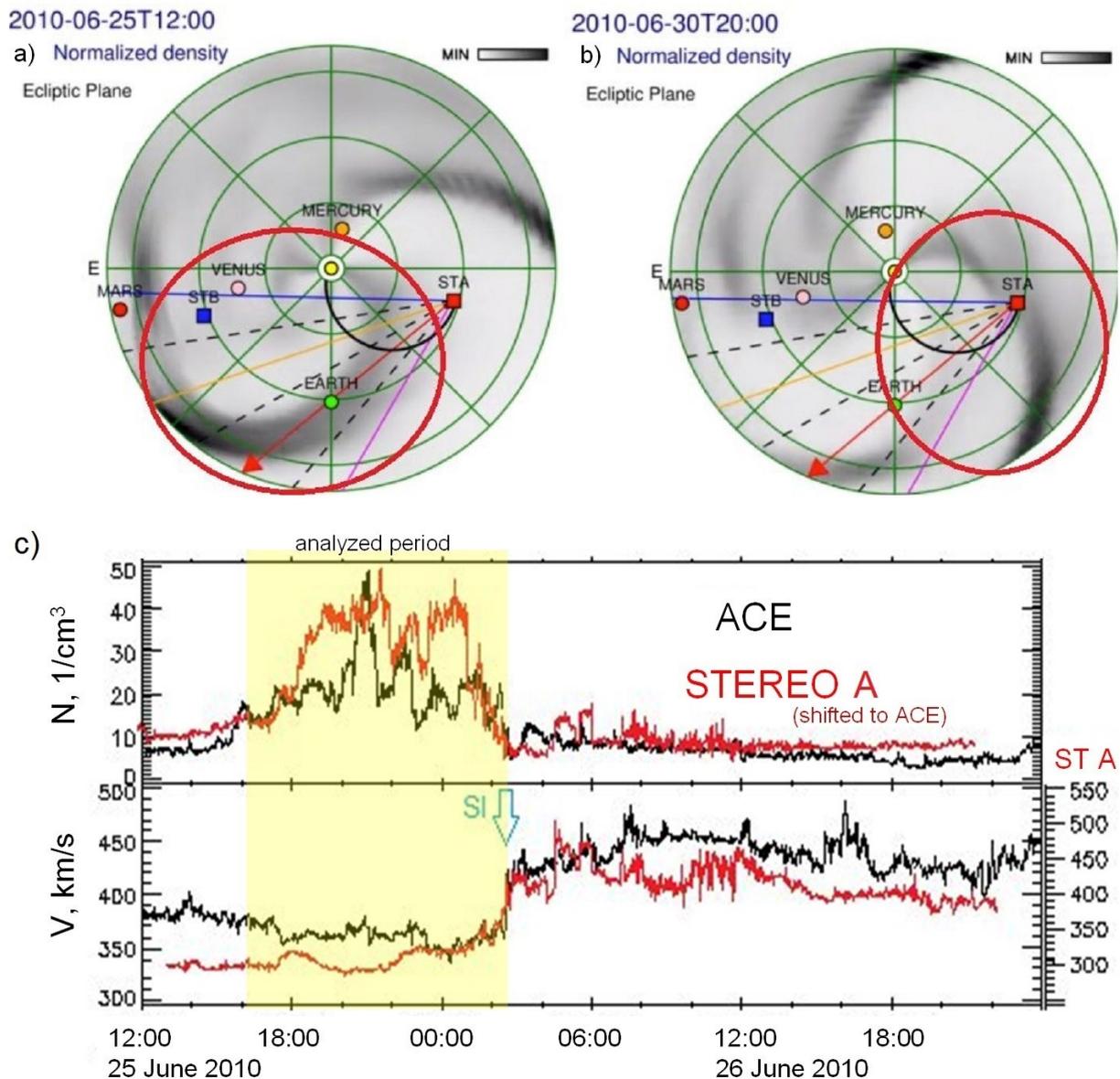

Figure 3. SIR subsequently observed by ACE and STEREO A on 25-30 June 2010. a) Normalized plasma density shown in the ecliptic plane is reconstructed by ENLIL predicting the HI STEREO observations. Earth – the green dot; the Sun is the yellow dot in the center. Red circles encompass the SIR of interest. SIR passes the Earth. b) The same but the SIR reaches STEREO A. Corresponding movie can be seen at http://helioweather.net/archive/2010/06/sta1dej.html . c) In situ observations of the solar wind density $N$ and speed $V$ by ACE (black) and STEREO A (red) are used to find the time of arrival of the stream interface to both spacecraft in order to compute the angular speed of the rotating SIR. The STEREO A profile is shifted to ACE for 24:41:30.

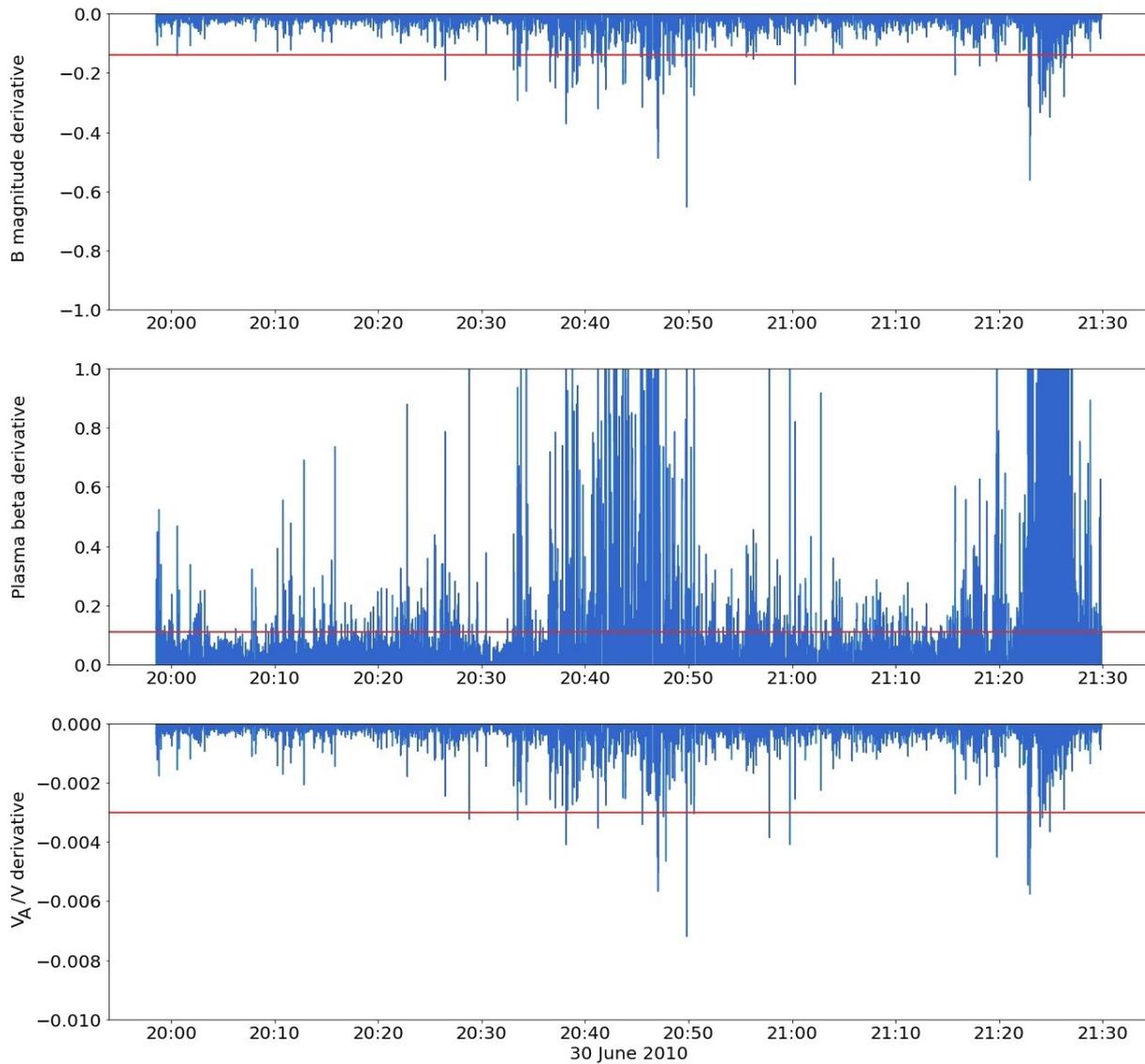

Figure 4. Example of automated identification of CSs via the three-parameter method. Based on the analysis of the ACE data, one second cadence. From top to bottom: derivatives of *B*, *β*, and $V_A/V$. Red lines indicate the noise cut off level (see 2.2. Method). Spikes indicate the CS location. Database of CSs identified at 1 AU is provided at https://csdb.izmiran.ru .

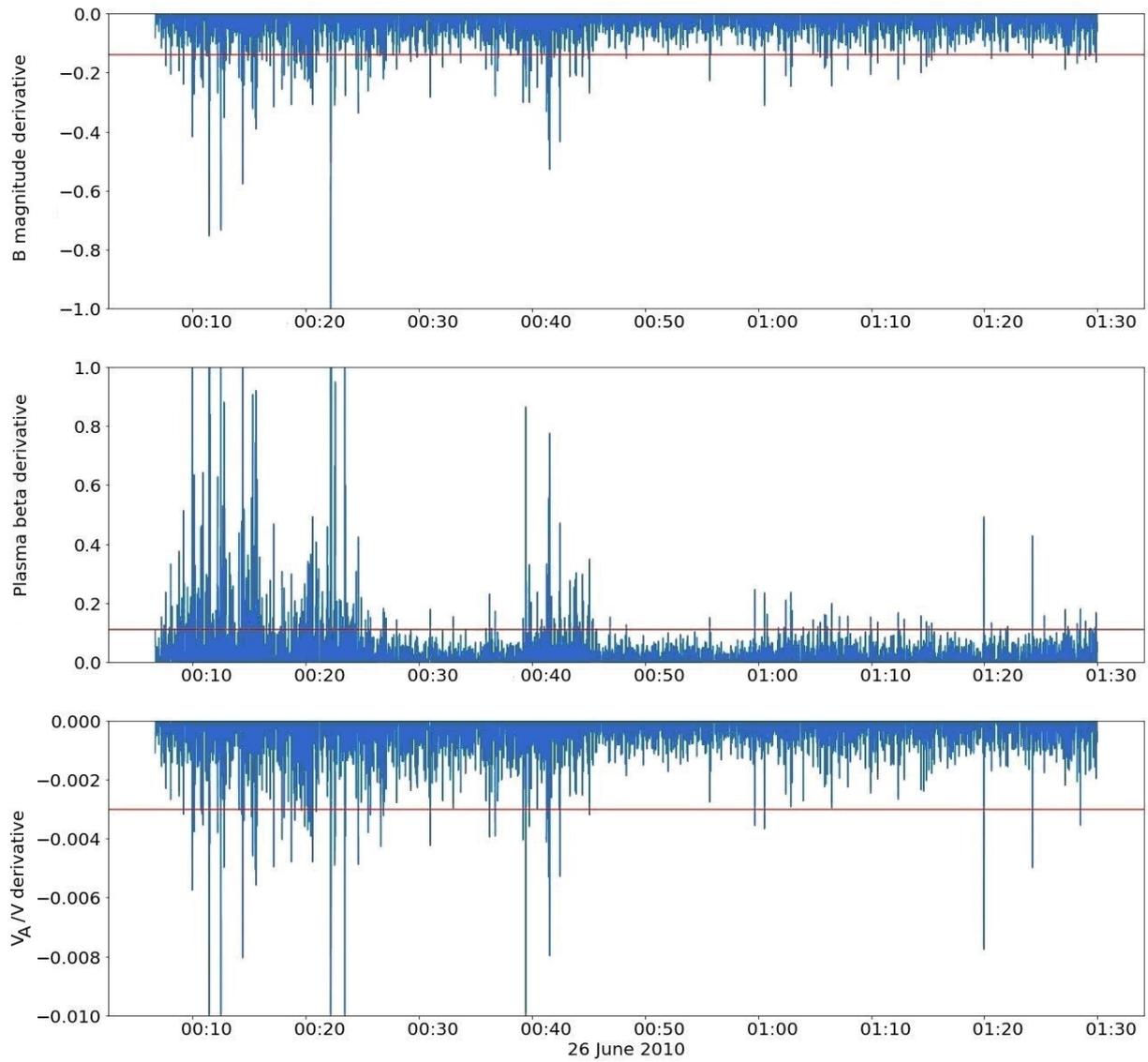

Figure 5. Analogous to Figure 3, but for STEREO A data, one second cadence.

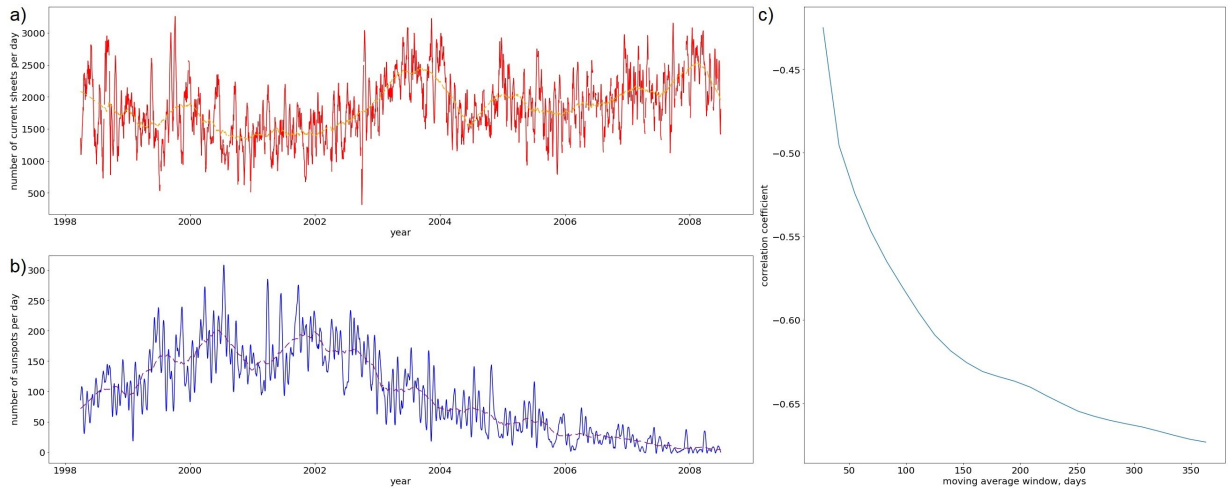

Figure 6. Number of CSs per day (*R*) observed at 1 AU in 1998-2010 vs the sunspot number. a) Red curve represents *R* smoothed with a 27-day (point) Savitzky-Golay filter, and the yellow curve shows a one-year smooth by the same method. The 3rd degree polynomial is used. b) Analogous to a) but for the number of sunspots per day. c) Correlation coefficient between a) and b) at different window widths.

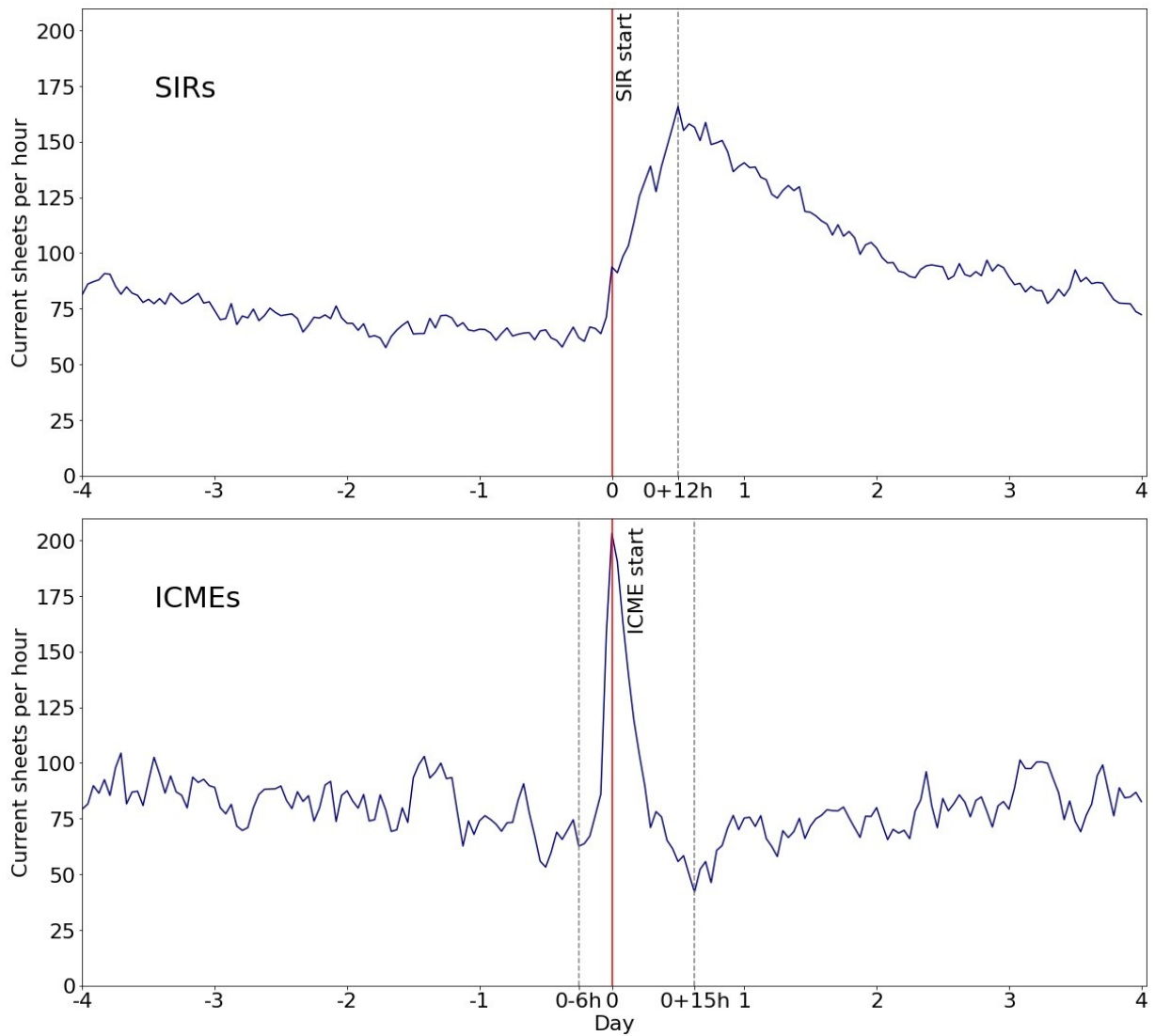

Figure 7. Results of the superposed epoch analysis of the number of current sheets per hour observed within and near SIRs (the upper panel) and ICME sheaths (the bottom panel) for 1998-2010. The leading edge of a SIR/ICME sheath is shown by the vertical red line (zero in the bottom panel).

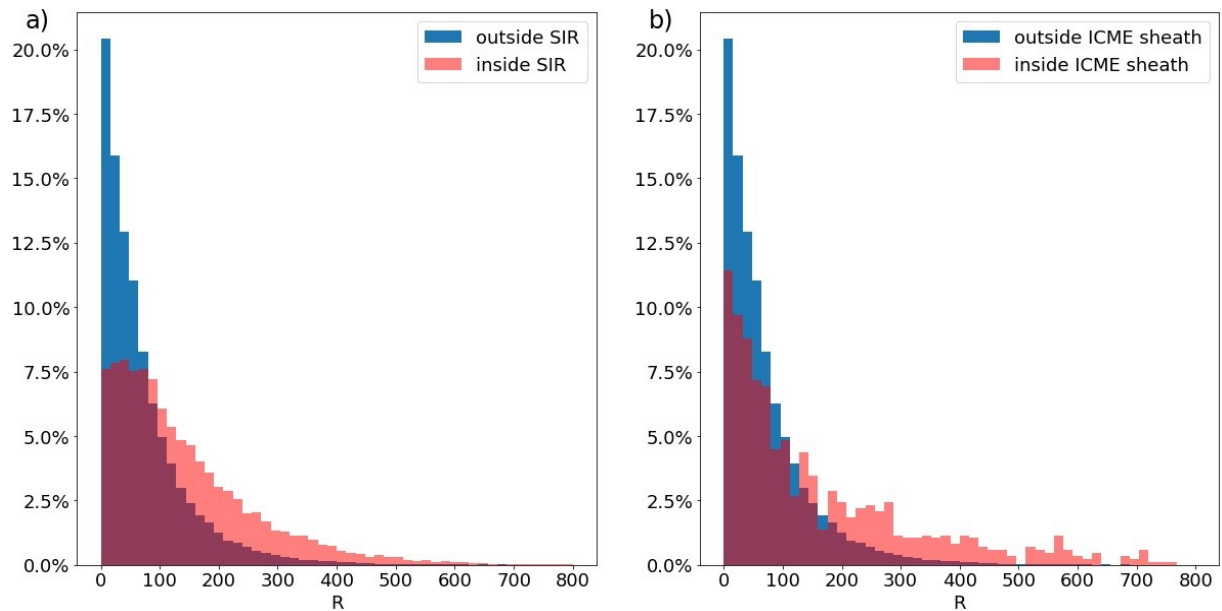

Figure 8. Distributions of *R* within and outside SIRs and ICME sheaths for 1998-2010. a) Histograms of the *R* occurrence calculated for hours in which SIRs are observed (red bars) and not observed (blue bars). b) Analogous to a) but for ICME sheaths. The occurrence of *R* is normalized and shown in percentages from the total number of events. Hours with SIRs are not included in the "no ICMEs" statistics, and hours during which ICME sheaths are observed are not included in the "no SIR" statistics not to contaminate the blue histograms. Hours analyzed: ICME sheath – 866, SIRs – 15540, calm – 68021. Red histograms are shifted to larger values. This indicates that R statistically increases within ICME sheaths and SIRs.

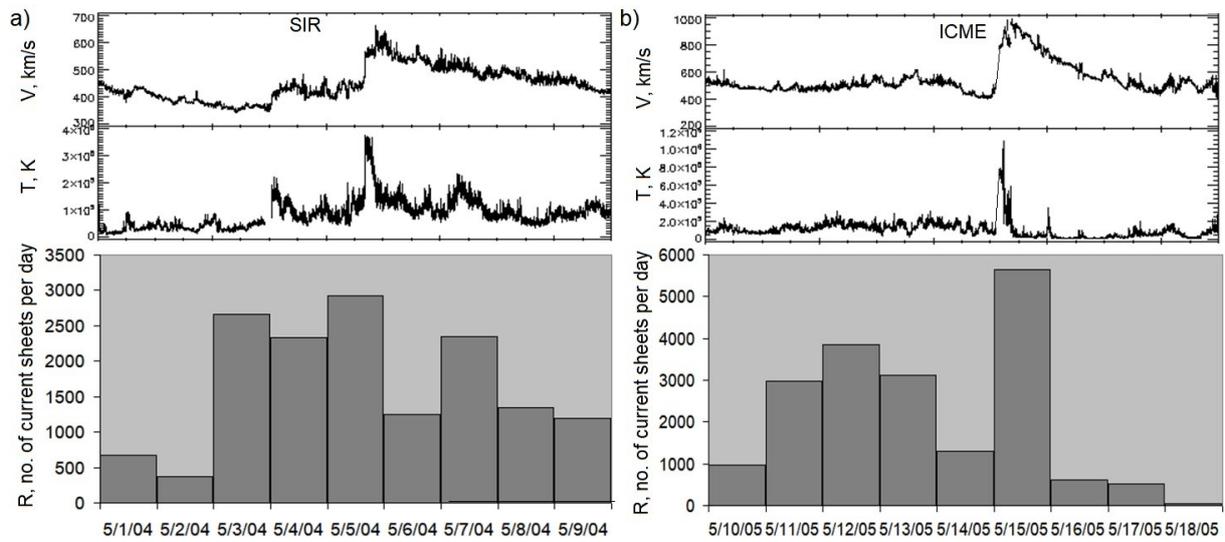

Figure 9. Typical variations of *R* associated with SIRs (a) and ICMEs (b). From top to bottom: solar wind speed *V*, temperature *T*, and *R* (from ACE).

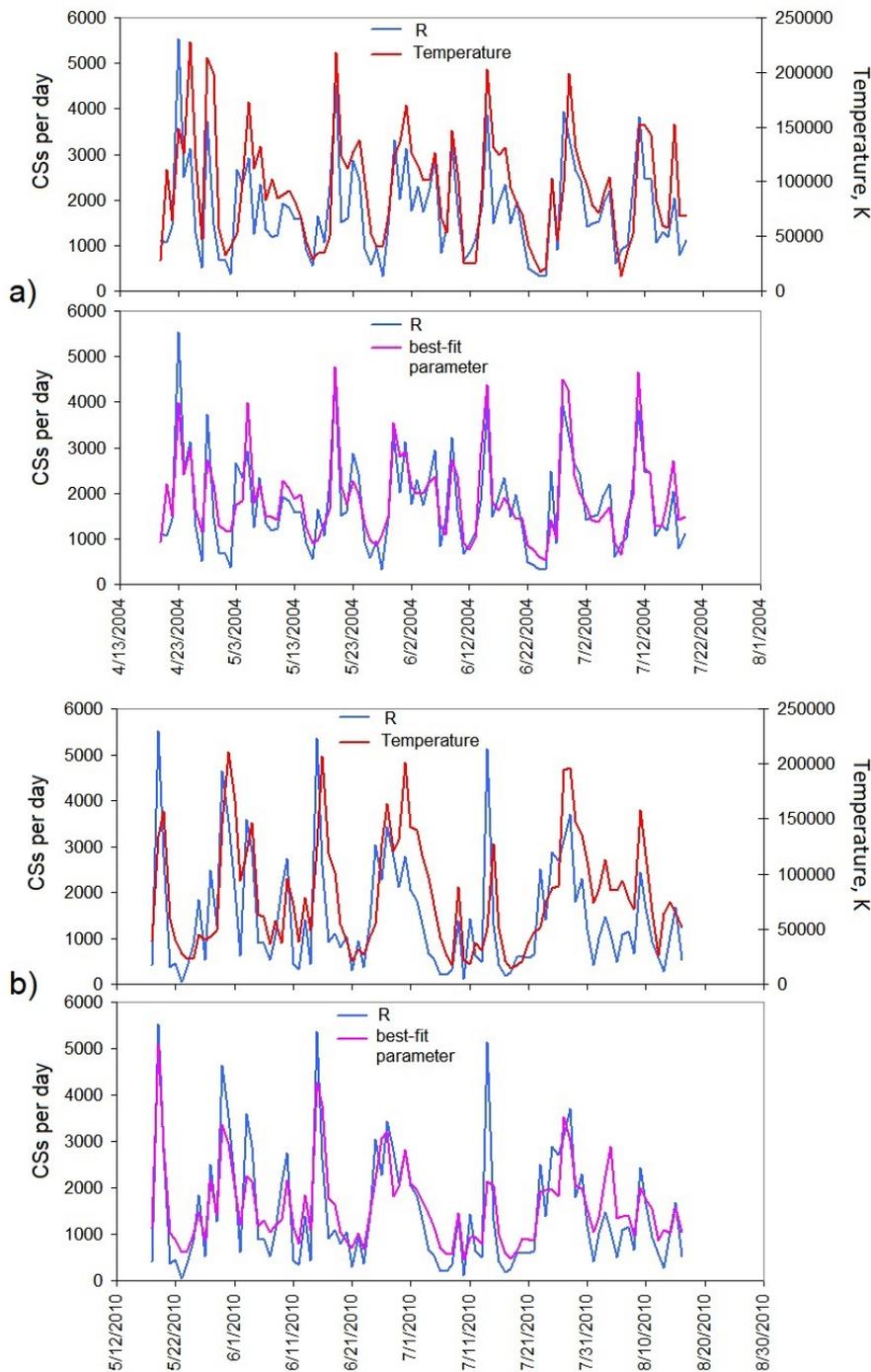

Figure 10. Finding the function that determines $R$. Examples are given for the three month-length periods in 2004 (a) and 2010 (b). Upper panels: the solar wind temperature $T$ vs $R$. Lower panels: $R$ vs the best-fit parameter $[V^2(N+5N2)+10T(N+N2)]/5000$, where $N2=2\text{cm}^{-3}$ is the level of the background density of the undisturbed solar wind. Correlation coefficients between $R$ and $T$ (upper panels) are 0.72 (a) and 0.64 (b). Correlation coefficients between $R$ and the best-fit parameter (lower panels) are 0.87 and 0.86, respectively. Based on the analysis of the data from ACE.

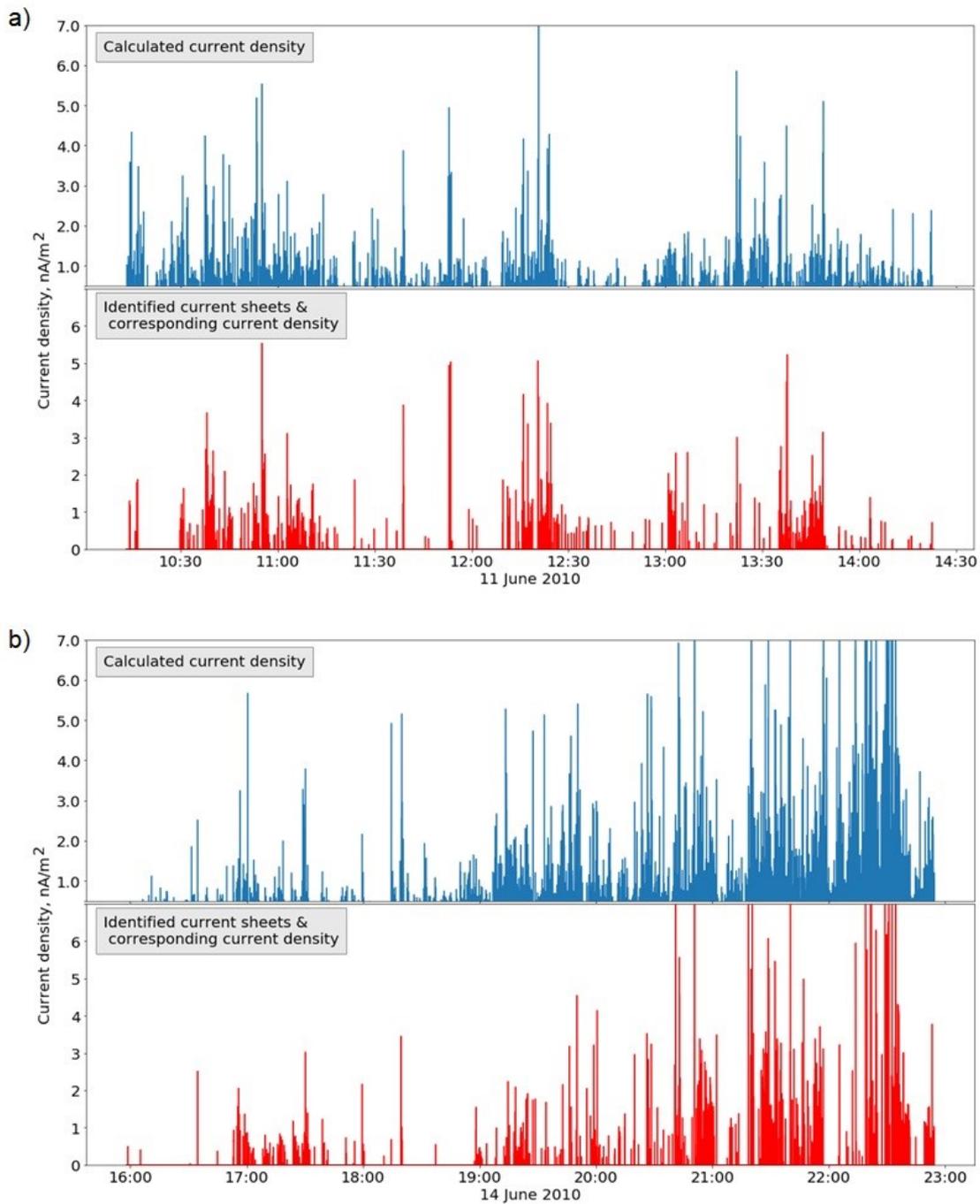

Figure 11. Electric current density calculated with an alternative method (top panels, blue bars) and CSs identified via the three-parameter method discussed above (bottom panels, red bars) for two randomly chosen time intervals of the event shown in Figure 2 by yellow stripe. Current density is cut off at 0.5 nA/m$^2$ to remove the noise. a) Example for the ACE spacecraft. b) Example for the STEREO A spacecraft.

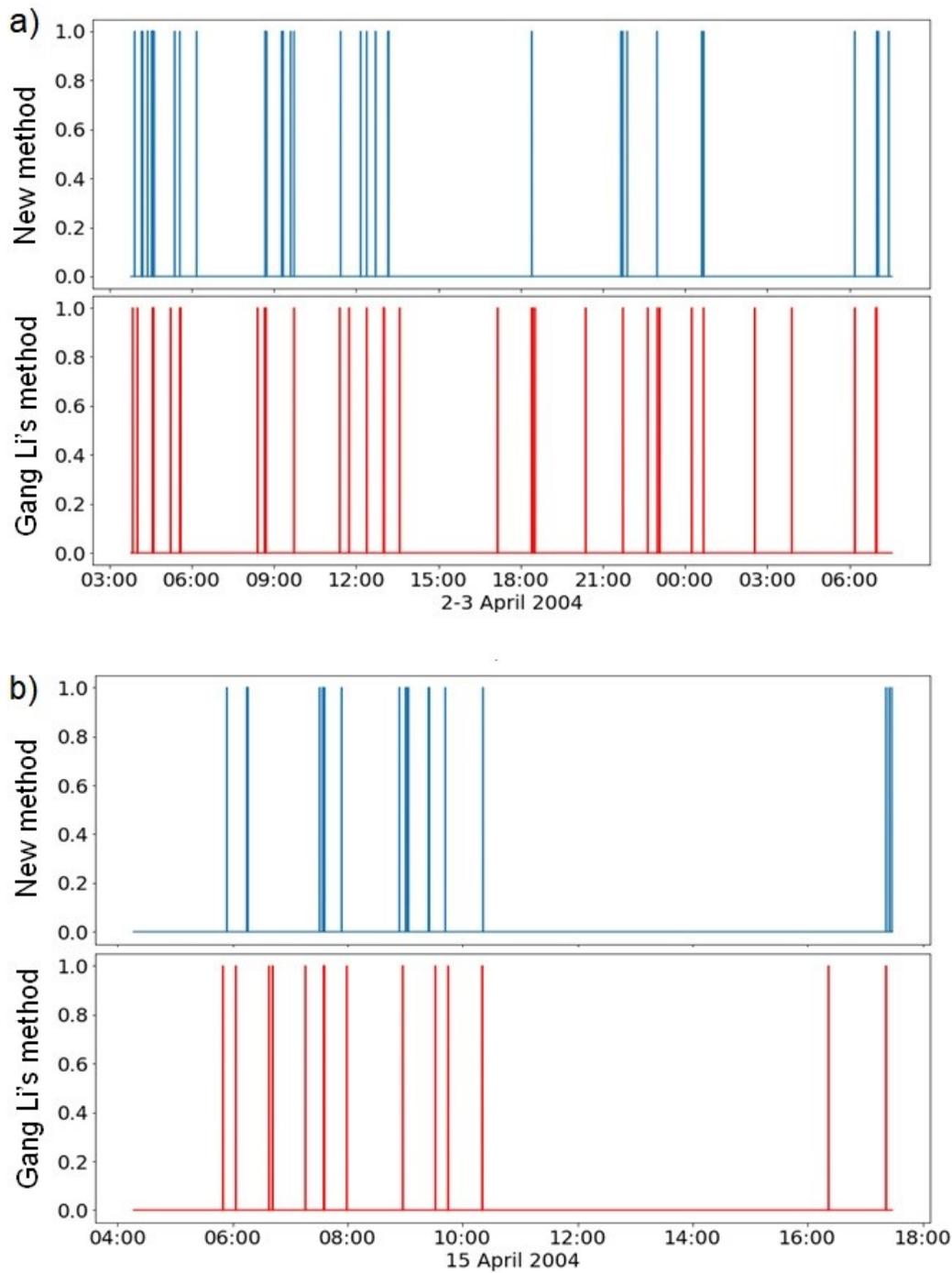

Figure 12. Comparison of results of the new three-parameter method (upper panels, blue bars) and Gang Li's method of CS identifying (lower panels, red bars). Randomly chosen time intervals on 2-3 April 2004 and 15 April 2004. One second cadence. Based on the analysis of the ACE spacecraft data.